\documentclass[12pt]{article}
\usepackage{graphicx}
\usepackage{titlesec}
\pdfoutput=1 
\usepackage[hyperfootnotes=true]{hyperref}
\hypersetup{colorlinks, citecolor=blue, linkcolor=blue, urlcolor=blue}
\usepackage{makecell}

\usepackage{hyperref}%
\usepackage{color,soul}
\usepackage{amsmath}
\usepackage{algorithm}
\usepackage{algorithmic}
\usepackage{changes}
\providecommand{\keywords}[1]
{\small	
\textbf{\textit{Keywords---}} #1}
\usepackage{geometry}
\graphicspath{{figs/}}
\usepackage{caption}
\usepackage{authblk}

\usepackage{caption}    
\usepackage[numbers]{natbib}

\usepackage{graphicx}
\usepackage{amsmath}
\usepackage{eqns}  
\usepackage{xcolor}

\title{\textbf{History-Matching of Imbibition Flow in Multiscale Fractured Porous Media Using Physics-Informed Neural Networks (PINNs)}}

\author[1]{Jassem Abbasi\thanks{Corresponding author: Jassem Abbasi (jassem.abbasi@uis.no)}}
\author[2]{Ben Moseley}
\author[3]{Takeshi Kurotori}
\author[4]{Ameya D. Jagtap}
\author[3]{Anthony R. Kovscek}
\author[1]{Aksel Hiorth}
\author[1]{Pål Østebø Andersen}

\affil[1]{\textit{Department of Energy Resources, University of Stavanger, Stavanger, Norway}}
\affil[2]{\textit{Seminar for Applied Mathematics, ETH Zürich, Zürich, Switzerland}}
\affil[3]{\textit{Department of Energy Science and Engineering, Stanford University, Stanford, USA}}
\affil[4]{\textit{Aerospace Engineering Department, Worcester Polytechnic Institute, Worcester, MA 01609, USA}}

\begin{document}
\maketitle

\begin{abstract}
In this work, we propose a workflow based on physics-informed neural networks (PINNs) to model multiphase fluid flow in fractured porous media. After validating the workflow in forward and inverse modeling of a synthetic problem of flow in fractured porous media, we applied it to a real experimental dataset in which brine is injected at a constant pressure drop into a CO$_2$ saturated naturally fractured shale core plug. The exact spatial positions of natural fractures and the dynamic in-situ distribution of fluids were imaged using a CT-scan setup. 
To model the targeted system, we followed a domain decomposition approach for matrix and fractures and a multi-network architecture for the separate calculation of water saturation and pressure. The flow equations in the matrix, fractures and interplay between them were solved during training. Prior to fully-coupled simulations, we suggested pre-training the model. This aided in a more efficient and successful training of the coupled system.
Both for the synthetic and experimental inverse problems, we determined flow parameters within the matrix and the fractures. Multiple random initializations of network and system parameters were performed to assess the uncertainty and uniqueness of the resulting calculations. The results confirmed the precision of the inverse calculated parameters in retrieving the main flow characteristics of the system. The consideration of multiscale matrix-fracture impacts is commonly overlooked in existing workflows. Accounting for them led to several orders of magnitude variations in the calculated flow properties compared to not accounting for them. To the best of our knowledge, the proposed PINNs-based workflow is the first to offer a reliable and computationally efficient solution for inverse modeling of multiphase flow in fractured porous media, achieved through history-matching noisy and multi-fidelity experimental measurements.

\end{abstract}


\section{Introduction}

Interpretation of multiphase flow phenomena in fractured porous media is challenging, because of their intrinsic multiscale nature \cite{Fung1991SimulationReservoirs}. In such environments, highly localized flow phenomena occur around fractures, which invalidate the assumptions behind traditional homogenization techniques \cite{Geiger2013AReservoirs}. As a result, the physical processes at the microscale must be explicitly modeled and resolved to capture the overall system behavior accurately at the coarse scale \citep{Mehmani2021StrivingLearned}. Almost all natural porous media are to some degree fractured; however, an inaccurate mathematical description or modeling of their influence on flow phenomena can lead to substantial errors  \citep{Berre2019FlowApproaches}. These errors are a result of an inaccurate handling of microscale force balances. For applications such as geological carbon storage, fluid flow in aquifers, and recovery of hydrocarbons  \citep{Rangel-German2006Multiphase-flowMedia} an inaccurate description of multiscale systems this can lead to incorrect predictions,  and suboptimal or even wrong decisions. Still, the most common approaches for estimating multiphase flow properties in porous media are developed for homogeneous systems with simplified frontal behavior \cite{McPhee2015CoreGuide}, but these methods are inapplicable for complex multiscale problems \cite{Yang2021RecentMedia}.  Experimentally, multiphase flow properties are characterized via parameter inversion, through history matching of the experimental measurements \cite{Finsterle1998MultiphaseOverview, Schembre2006EstimationExperiments}. In many cases, the multiscale problem is bypassed by selecting rock samples that are relatively homogeneous or making different simplifying assumptions \citep{Berre2019FlowApproaches}. However, when boundary conditions change, experimental simplifications can lead to unreliable parameter estimations, particularly for formations such as shale, where complex fracture networks significantly influence fluid flow behaviour \citep{Salem2022ImpactShale}.

Furthermore, recent advancements in measurement and imaging technologies have revolutionized our ability to characterize complex, multiphase, flow patterns in porous media. These improvements significantly enhance our capacity to investigate multiphase flow dynamics \citep{Tekseth20244DTomography}. The coupled matrix and fracture multiphase flow systems require a multiscale description and must be treated carefully when solved numerically \citep{Odster2019AMedia}. Fractures introduce discontinuities in the flow field, which can lead to numerical instabilities and reduced accuracy \cite{Su2015MultiphaseFractures}. Additionally, a multiscale method requires a fine discretization of the domain to capture both multiphase discontinuities and fine scale behavior, which results in high computational costs \citep{Bastian2000NumericalMedia}. Typically, one uses numerical methods based on finite-element and finite-volume to solve the problem more efficiently, due to the capability of these methods in handling complex fracture geometries and incorporating discontinuities \citep{Komijani2023AnCoupling, Myner2016AGrids}. However, the implementation of the algorithms is complex, and the numerical schemes typically faces challenges such as high computational requirements and in some cases low convergence reliability \citep{Yang2021RecentMedia}, which could make them impractical and time-consuming for inverse calculation purposes. Given these limitations and opportunities, there is a demand to develop more precise and computationally reliable mathematical tools, especially for the inverse interpretation of multi-fidelity datasets.

In recent years, Physics-Informed Neural Networks (PINNs) have emerged as a mathematical technique, seamlessly integrating high-fidelity and noisy datasets and domain expertise to solve differential equations governing various engineering and scientific problems \citep{Raissi2019Physics-informedEquations}. More specifically, it has shown potential to solve problems of multiphase flow in porous media in forward or inverse modes \citep{Latrach2024ASystems}, e.g., modelling of countercurrent spontaneous imbibition \citep{Abbasi2023SimulationNetworks}, solving multiphysics coupled flow problems\citep{Amini2023InverseNetworks} and modelling density-driven flow in porous media for CO$_2$ storage \citep{Du2023ModelingSequestration}. However, current studies are primarily focused on systems with homogeneous properties, often overlooking systems with multiscale features. The application of PINNs to multiscale problems is an emerging area of research, with some studies proposing various techniques in recent years \citep{Alber2019IntegratingSciences}. 
For instance, \citet{Wang2021OnNetworks}, in the case of wave propagation and the Poisson equation, proposed to use Fourier feature embedding to capture the behavior of the system at various spatial scales and at the same time avoiding the potential spectral biases in the PINNs predictions. Spectral bias is defined as the tendency of networks to learn solutions with the lowest possible frequencies. \citet{Weng2022MultiscaleKinetics} proposed multiscale PINNs to model chemical kinetics that happen at very different time scales. They utilized different NNs with similar architectures for the different  reaction rates. \citet{Riganti2023AuxiliaryProblems} applied multi-output PINNs with multiple auxiliary variables associated and the Legendre terms to solve the Boltzmann equation efficiently. \citet{Moseley2023FiniteEquations} introduced Finite-Basis PINNs, by decomposing the domains of the problems, e.g., (2+1)-dimensional wave propagation equations, using a finite set of basis functions, inspired by finite element methods. However, the literature shows significant gaps in the application of PINNs for addressing forward modeling and inverse parameter identification in multiphase flow in naturally fractured porous media, which has broad implications for geosciences.

In this work, we have implemented a workflow based on the multiscale PINNs architecture for the forward and inverse simulation of (3+1) dimensional multiphase flow in fractured porous media. To the best of our knowledge, the proposed workflow is the first to offer a PINNs-based approach for reliable and computationally efficient inverse modeling of multiphase flow in fractured porous media, suitable for history-matching noisy and multifidelity experimental measurements. The work demonstrates clear advantages in both accuracy and computational speed for inverse problems. Importantly, it effectively addresses the complexity inherent in modeling multiphase flow within fractured porous media, effectively capturing multifidelity observational data. The proposed workflow can be a critical factor for decision-making in industrial applications, including geological carbon storage and optimized reservoir management.

The proposed workflow utilizes a multi-network architecture combined with various regularization techniques, such as data resampling during training, along with a step-wise training strategy to efficiently achieve the desired solutions. The robustness of the computational framework was first demonstrated through validation using a synthetic benchmark problem, and the performance compared against a finite-difference (FD) based numerical simulator, as a common approach in the community. Then we investigated a complex experimental scenario of water imbibition in CO$_2$ saturated fractured shale rock \citep{Kurotori2023MixedMedia}. We incorporated a large multi-fidelity observational dataset, including 3D in situ saturation data measured using a high-resolution computed tomography (CT) scan setup, in the process of inverse calculations. In this case, we applied the workflow for the simultaneous extraction of flow properties of the matrix and fracture system.


\section{Methods}
\label{sec:methods}

\subsection{Experimental Data}
\label{sec:experimentaldata}
\begin{figure}[h]
\begin{center}
\includegraphics[width=.999\textwidth]%
    {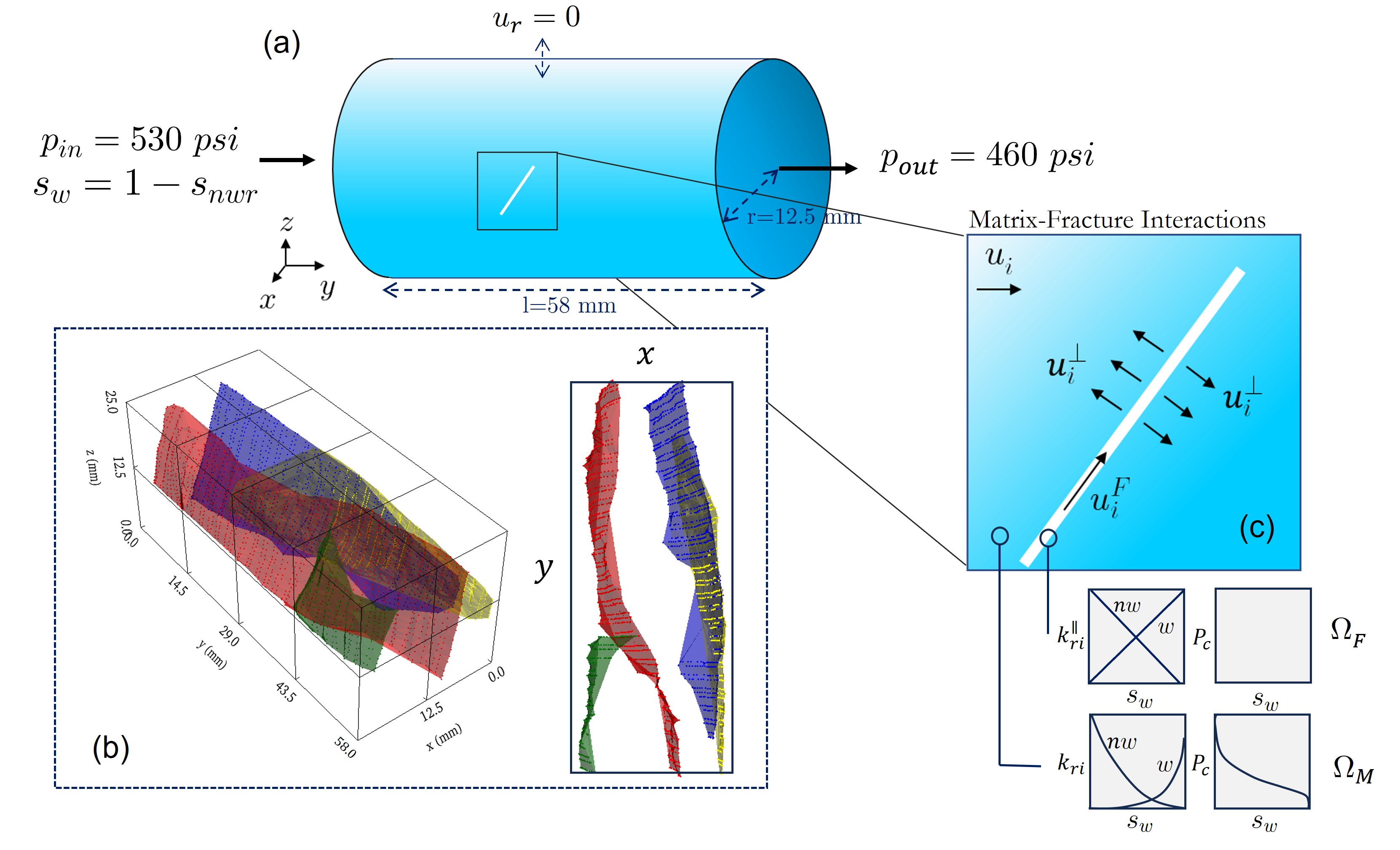}
\end{center}
\caption{\textbf{A schematic of the core and the applied boundary conditions, as well as the distribution of fractures in the core.} \textbf{a)} The simplified geometry of the matrix-fracture system and the impacting boundary conditions, \textbf{c)} The spatial distribution of the matrix and fracture collocation points, in three-dimensions (left), and x-y projection (right), \textbf{c)} the mathematical notations in fractures}
\label{fig:coreschematic}
\end{figure}

This study is motivated by the need for a workflow that efficiently can evaluate and history-match an experimental dataset injecting brine into a  CO\textsubscript{2}-saturated naturally fractured shale rock core, as detailed in \citet{Kurotori2023MixedMedia}. 
Shales are considered tight media with very low permeability and have only the recent decades been developed for hydrocarbon production due to advances in hydraulic fracturing and drilling technology \citep{Alexander2011ShaleRevolution}. The affinity of shale to adsorb CO$_2$ (e.g. over CH$_4$, shale gas) has also resulted in opportunities to sequester carbon in shales \citep{Klewiah2020ReviewShales,Godec2014EnhancedPotential}.
The examined cylindrical Wolfcamp shale core had a length of 5.8 cm, diameters of 2.5 cm (see Fig. \ref{fig:coreschematic}a), an average permeability of \(1.97e-17\) m$^2$ (0.020 mD) and porosity of 0.102 (see Fig. \ref{fig:2}a). The reported permeability represents the average permeability of both the matrix and fractures, with the matrix permeability expected to be significantly lower. The core was initially saturated with CO$_2$, then brine was injected at fixed pressure drop (530 psi injection pressure and 460 psi back-pressure) for 311 hours while confined at 700 psi. Dynamic computed tomography (CT) images were captured at intervals, recording in-situ saturation, phase recovery, and injected fluid volume over time. Different fractures were identified from the CT scan images \cite{Murugesu2024CoupledCaprocks}, which the extracted coordinates are shown in Fig. \ref{fig:coreschematic}b. History matching and inverse calculations in this study relied on several key experimental measurements (also see Fig. \ref{fig:colls}e):

\begin{itemize}
\item\textbf{Recovery Factor (RF)}: The volume fraction of the pore space displaced by the injected fluid over time. We calculated the RF curve by averaging the saturation of the wetting phase fluid obtained from CT scan images.
\item\textbf{Boundary Conditions}: The constant pressures at the inlet and outlet boundaries were known based on the experimental setup yielding a fixed pressure drop over the core.
\item\textbf{Injection Volume}: The cumulative volume of water injected as function of time was measured.
\item\textbf{Spatial Saturation Data}: Spatial distributions of fluid saturation within the core sample at different times, obtained through CT scans (19 snapshots).
\end{itemize}

Regarding the fluid properties, by considering the laboratory conditions, the viscosities of water  and CO$_2$  are assumed to be 0.89 cP \citep{Haynes2016CRCPhysics} and 0.0157 cP \citep{Fenghour1998TheDioxide}, respectively, with their interfacial tension (IFT) assumed to be 0.04 \(N/m\).  \citep{Shiga2023InterfacialTemperature}. Water and  CO$_2$ density also were set to be 998.7 and 78.9 \( kg/m^{3}\), respectively \citep{Houben2021SupercriticalMembranes}.

\subsection{Mathematics of Flow in Fractured Porous Media}
\label{sec:mathematicsofphysics}

To develop the stated workflow, we approach the problem of 3D multiphase flow in a fractured porous media at core scale. 
We refer to the low-permeability non-fractured zone as matrix domain, and the high-permeability fractured zone as fracture domain. The model is inferred from that the matrix domain \((\Omega_M)\) occupies the majority of the pore volume and that the fracture domain \((\Omega_F)\) has a much higher permeability than the matrix.
 A simplified schematic of the sample set-up is visualized in Fig. \ref{fig:coreschematic}a. The complete information regarding the geometric structures of the core and the fractures are given in \citet{Kurotori2023MixedMedia}.

The core is placed horizontally, and the fractures are distributed in the core with different orientations. The radial surface of the core at \(r_c\) is closed (no-flow boundary). The inlet and outlet faces are at \(y=0\)  and \(y=L\), respectively (Fig. \ref{fig:coreschematic}a).  Water and CO\textsubscript{2} are considered to be the wetting \((w)\) and non-wetting \((nw)\) phases, respectively. In line with the experimental conditions (the pressures were high and the water was equilibrated), we neglected both the compressibility and the solubility of the phases. The system is initially saturated with the \(nw\) phase. Also, the effects of gravity have been ignored due to low values of the Bond number (ratio of gravity to capillary forces), i.e., $Bo << 1$ \cite{Li2018DynamicsMedia}. Furthermore, the system, including the rock and fluid properties were considered incompressible, due to the small pressure and temperature variations in the system.

The definition of the system as a fracture-matrix system implies that the flow phenomena are governed by mixed-imbibition mechanisms, where the fracture flow is driven by forced imbibition, and the matrix flow is governed by spontaneous imbibition between the fracture(s) and the matrix \cite{Kurotori2023MixedMedia}. Forced imbibition is the case in which an external force drives the wetting phase fluid through the porous media, overcoming any resistance from the medium and the nonwetting phase. Spontaneous imbibition, on the other hand, is the process in which the wetting phase fluid flows into a porous media via capillary forces without the need for external pressure \citep{Anderson1987WettabilityPressure}. Especially in shales, microfractures can be abundant while the micro- and nanopores in the matrix can produce very high capillary pressures driving spontaneous imbibition \cite{Alipour2022EmpiricalShale, Roychaudhuri2013AnShales}. In the following, the mathematics of flow in matrix and fracture domains are discussed.

\subparagraph{Matrix flow. }
The Darcy velocity (\(u_i\)) of each phase is determined by Darcy's law
\begin{equation}
u_i =-\lambda_i \nabla p_i,\ \ \lambda_i=\frac{Kk_{ri}}{\mu_i},\ \ \left(i=w,nw\right)\label{eq:darcy}
\end{equation}
where \(\lambda_i\) is phase mobility, \(p_i\) is the phase pressure, \(K\) and \(k_{ri}\) are absolute permeability and relative permeability, respectively, and \(\mu_i\) is the phase viscosity. The index \(i\) refers to the \((w)\), and \((nw)\) phases. The capillary pressure, \(p_c = p_{nw}-p_{w}\), relates the phase pressures.  The conservation law for phase mass transport in the matrix zone is defined as
\begin{align}
\phi\partial_t\left(\rho_is_i\right)=-\nabla\left(\rho_iu_i\right),\ \ \left(i=w,nw\right),\ \ {\in\Omega}_M \label{eq:matrix_cont}
\end{align}
We define the initial conditions by zero water saturation and an initial pressure (of the CO$_2$ phase) as
\begin{align}
\left\{
\begin{array}{ll}
{s}_{w}(x,y,z,t = 0) = 0 \\
{p}_{nw}(x,y,z,t = 0) = {p}_{i}
\end{array} 
\right. \ \ \ {, \in \Omega}_M
\label{eq:eqICmatrix}
\end{align}

At the inlet face ($y=0$), the wetting phase is injected at constant pressure ${p}_{in}$. At this face, it is assumed the wetting phase takes the highest mobile saturation:
\begin{align}
\left\{
\begin{array}{ll}
{s}_{w}(x,y = 0,z,t) = 1 - {s}_{nwr} \\
{p}_{w}(x,y = 0,z,t) = {p}_{in}
\end{array} 
\right.  \ \ \ {, \in \partial \Omega}_M
\label{eq:eqBC0matrix}
\end{align}
where $s_{nwr}$ is the residual saturation (where it becomes immobile) of non-wetting phase.
The production from the outlet face \((y=L)\) occurs at a constant pressure, \({p}_{out}\)
\begin{align}
{p}_{nw}(x,y = L,z,t) = {p}_{out}  \ \ \ {, \in \partial \Omega}_M
\label{eq:eqBC1matrix}
\end{align}
At the outlet face of the core, no explicit saturation constraints are imposed. The radial faces \((r = r_c)\) are sealed from flow in radial directions
\begin{align}
 \frac{\partial p}{\partial r} |_{r = r_c} = 0 \ \ \ {, \in \partial \Omega}_M 
\label{eq:eqBCrmatrix}
\end{align}

The included matrix flow mechanisms state that the flow in the matrix can be governed by viscous and capillary forces while gravity effects are assumed to be negligible.

\subparagraph{Fracture flow. }
Fractures are characterized as narrow, high permeability zones enclosed by less permeable matrix. The aperture between the fracture surfaces is generally small (e.g. micrometers) and can be occupied by materials, that affect its effective porosity and permeability \citep{Wu2016MultiphaseMedia}. 

Although fractures are discontinuities in the porous medium, they are considered as separate porous domains \((\Omega_F)\) with high permeability (see Fig. \ref{fig:coreschematic}c). 
This makes the equations in matrix and fracture media consistent  \citep{Martin2005ModelingMedia}. The phase flux in a fracture is written by modification of Darcy's law:
\begin{equation}
u_i^{F}=-\lambda_i^{F} \nabla p_i ,\ \ \lambda_i^F=\frac{K^F k_{ri}^F}{\mu_i},\ \ \left(i=w,nw\right)\label{eq:darcy_frac}.
\end{equation}
\(K^F\) is the fracture permeability, and \(k_{ri}^F\) the fluid relative permeability in the fracture.

\subparagraph{Matrix-fracture interactions. }

Accurate treatment of multiphase flow in fractured porous media depends on the modeling of the dynamic matrix-fracture coupled interactions. Utilizing Darcy's law \cite{Berre2019FlowApproaches}, the matrix-fracture flux is approximated by 
\begin{align}
u_i^\bot=-K^\bot\lambda_i^\bot\left[\frac{ p_{M,i}-p_{F,i} }{\tfrac{e_V}{2}}\right],\ \ \lambda_i^\bot \simeq \frac{k_{ri}}{\mu_i},\ \ \left(i=w,nw\right), \label{eq:e8}
\end{align}
where \(K^\bot\) refers to the matrix-fracture permeability normal to the fracture surface, and \(e_V\) is the fracture average aperture. Here, we assumed the matrix-fracture interactions are mainly controlled by the matrix permeability, then \(K^\bot = K \).  In this study, \(e_v\), is assumed to be 0.001 m for all the fractures, and is divided by two to take the pressure difference from the center of the fracture. We can rewrite equation (\ref{eq:e8}) to obtain the mass transfer rate [kg/m\(^3\)/s] between the two media as
\begin{align}
q_i^\bot=\frac{2\rho_i}{e_V} u_i^\bot ,\ \ \left(i=w,nw\right). \label{eq:massrate}
\end{align}
The factor 2 denotes the flow across both sides of fracture to matrix, assuming the equal flow characteristics across the two surfaces of the fracture. 
The equation of mass conservation in the fracture is similar to that in the matrix (equation (\ref{eq:matrix_cont})), but with the additional source term \( q_i^\bot \): 
\begin{align}
\phi^F\partial_t\left(\rho_is_{i}^{F}\right)=-\nabla \cdot\left(\rho_iu_i^F\right)\ -q_i^\bot,\ \ \left(i=w,nw\right),\ \ {\in\Omega}_F \label{eq:massbalF}
\end{align}
At the contact point of the matrix and fractures, the continuity equation should also be solved in the matrix domain. So, we may rewrite equation (\ref{eq:matrix_cont}) as 
\begin{align}
\phi\partial_t\left(\rho_is_i\right)=-\nabla\cdot\left(\rho_iu_i\right)+q_i^\bot,\ \ \left(i=w,nw\right),\ \ {\in\Omega}_M \label{eq:matrix_frac_cont}
\end{align}

\subparagraph{Closure equations. }
Capillary pressure and relative permeability curves are crucial input to model multiphase flow in porous media, e.g., in equations (\ref{eq:matrix_cont}), (\ref{eq:massbalF}) and (\ref{eq:matrix_frac_cont}). These curves are normally described by correlations with a few tuning parameters. In this work, we apply an extended version of the Corey function \citep{Corey1954ThePermeabilities} to model relative permeability curves:
\begin{align}
{k}_{rw}({S}_{w}) = {k}_{rw}^{max}({S}_{w})^{{n}_{w}};\ \ \ \ \ \ {n}_{w} = {n}_{w1}{S}_{w} + {n}_{w2}(1 - {S}_{w}) \label{eq:krw_eq}
\end{align}
\begin{align}
{k}_{rnw}({S}_{w}) = {k}_{rnw}^{max}(1 - {S}_{w})^{{n}_{nw}};\ \ \ \ \ \ {n}_{nw} = {n}_{nw1}{S}_{w} + {n}_{nw2}(1 - {S}_{w})
\label{eq:krnw_eq}
\end{align}

\(n_w\) and \(n_{nw}\) are saturation exponents that vary linearly with \(S_w\) \cite{Andersen2023Early-andCoefficient} where \(S_w\) is the normalized water saturation calculated as:
\begin{align}
{S}_{w} = \frac{{s}_{w} - {s}_{wc}}{1 - {s}_{nwr} - {s}_{wc}};
\label{eq:Sweq}
\end{align}
The capillary pressure curve \(p_c(S_w)\) [Pa] is calculated via Leverett scaling \citep{Leverett1942Dimensional-modelBehavior}

\begin{align}
p_c = \sigma \sqrt{\frac{\phi}{K}} J
\label{eq:pceq}
\end{align}
In this context, J(S$_w$) represents the Leverett J-function, a dimensionless function that describes the shape of the capillary pressure curve. We define J-function via the modified \citet{Bentsen1977UsingCurve} correlation
\begin{align}
J(S_w) =  - {J}_{1}\mathrm{\ln}(\frac{{S}_{w}}{{S}_{eq}}) + {J}_{2}\mathrm{\ln}(\frac{1 - {S}_{w}}{1 - {S}_{eq}})
\label{eq:Jeq}
\end{align}

The equation is constrained so that \(J(S_{eq})=0\) and that \(dJ/dS_w\) becomes infinite at endpoints.

\subsection{Known and unknown parameters. }
The determination of fluid properties, as well as effective porosity and permeability of the rock sample can be done fairly independently using established methods. However, determination of the multiphase flow properties of the system - i.e., the relative permeability (\( k_r\)) and capillary pressure (\( p_c\)) of the matrix and fracture domains - is challenging but essential for accurately understanding the governing mechanisms. Measurement of these curves is challenging and typically requires indirect methods, such as history matching using numerical simulations \citep{Krause2015AccurateCurves}. This primarily requires multiphase experimental core flooding tests, in which the cylindrical rock sample is first saturated by a phase, then injected by the second phase at controlled pressures or flow rates. Established procedures for determining multiphase properties however do not account for the presence of complex geometries.  History matching in complex geometries, such as fractured porous media, is a challenging task that involves significant computational complexities. The challenge becomes more serious when high-fidelity datasets, such as CT-scan images, must be matched. In this work, we propose the application of PINNs as a reliable framework for history matching and characterization of these flow parameters. 
The matrix \( k_r\) curves can be described via seven parameters related to the Corey model (equations (\ref{eq:krw_eq}) and (\ref{eq:krnw_eq})): \({k}_{rw}^{max}\), \({k}_{rnw}^{max}\), \(n_{w1}\) and \(n_{w2}\), \(n_{nw1}\) and \(n_{nw2}\) and $s_{nwr}$. Also, the \( p_c\) curve can be characterized via \(J_{1}\), \(J_{2}\) and $\sigma$ values, based on equation (\ref{eq:pceq}). 

\subparagraph{Assumptions.} Although we had access to the average permeability of shale rock, an accurate measurement of matrix permeability ($K$) was not available. Instead, we assumed a constant value and conducted inverse calculations based on uncertain flow functions. Then, the calculated values of relative permeability are dependent on the assumed absolute permeability. However, the applied assumption does not affect the generality of either the effective flow properties of the porous media or the methodology used. Then, the fracture permeability (\(K_{F}\)), which significantly impacts the overall flow rate during the experiments, was estimated based on flow rates obtained from the given pressure drop during history-matching.  Furthermore, due to the small pore volume of the fractures, we assumed a constant fracture porosity, equal to the matrix porosity ($\phi^F=\phi$). We also assumed linear relative permeability curves and zero capillary pressure for fractures \citep{DeLaPorte2005TheReservoirs}. This means that, unlike in the matrix medium, the fracture flow properties are determined by assuming constant relative permeability values, while optimizing the absolute permeability.

\subsection{Methodology}

Physics-Informed Neural Networks (PINNs) were first applied by \citet{Raissi2019Physics-informedEquations} as an efficient method for solving differential equations by combining neural networks with physics-based principles. In this study, we defined neural networks (separately for matrix (\( \mathcal{N}_m \)) and fracture (\( \mathcal{N}_f \)) domains), to emulate the functional dependency between the independent and dependent variables
\begin{align}
{s}_{w},{p}_{nw} = \mathcal{N}^{M}(x,y,z,t,\theta,\eta)
\label{eq:pceq}
\end{align}
\begin{align}
{s}_{w}^{f},{p}_{nw}^{f} = \mathcal{N}^{F}(x_{f},y_{f},z_{f},t,\theta_{f},\eta)
\label{eq:p1}
\end{align}
where, [\(x, y, z\)] and [\(x_f, y_f, z_f\)]   are the spatial coordinates corresponding to the collocation points of matrix and fracture, respectively. The output of the networks are the state variables of the system, that is \(s_w\), and \(p_{nw}\). \( \theta\) and \( \theta_f\) are the network trainable parameters.
The goal is to solve an optimization problem where:
\begin{align}
\theta^{m,f},\eta = \arg\min(\mathcal{L}_{t})
\label{eq:p2}
\end{align}
where \( \eta\) is a set of trainable inverse parameters calculated during the inverse calculations. Also, \( \mathcal{L}_{t} \) is the total loss term, defined to minimize the errors in both the physical flow equations in the matrix and fracture, as well as the available observation data. 
\begin{align}
\mathcal{L}_{t} = \mathcal{L}_{t}^{M} + \mathcal{L}_{t}^{F} + \mathcal{L}_{t}^{D} 
\label{eq:p2}
\end{align}
where \(\mathcal{L}_{t}^{M}\) and \(\mathcal{L}_{t}^{F}\) are the total loss terms corresponding to the initial/boundary conditions, as well as the PDE residuals for matrix and fracture domains, respectively. Minimizing these terms ensures the PINNs solution respects the physical constraints within their domain.
Furthermore, \(\mathcal{L}_{t}^{D}\) represents the total loss term corresponding to the errors in the predictions of PINNs compared to the observational data. The complete explanation of the PINNs implementation and the underlying loss terms is provided in Appendix \ref{sec:PINNsimplementation}.

To address the inverse problem, we adopted an ansatz approach for representing the relative permeability and capillary pressure curves. This means a trainable vector ($\eta$) was defined, where each element corresponded to a parameter within the chosen correlations for these curves (equations (\ref{eq:krw_eq}), (\ref{eq:krnw_eq}) and (\ref{eq:pceq})). More information regarding the initialization and definition of the inverse variables is provided in Appendix \ref{sec:PINNsComputationalstrat}.

\subsection{Model Architecture}
\begin{figure}[h]
\begin{center}
\includegraphics[width=.999\textwidth]%
    {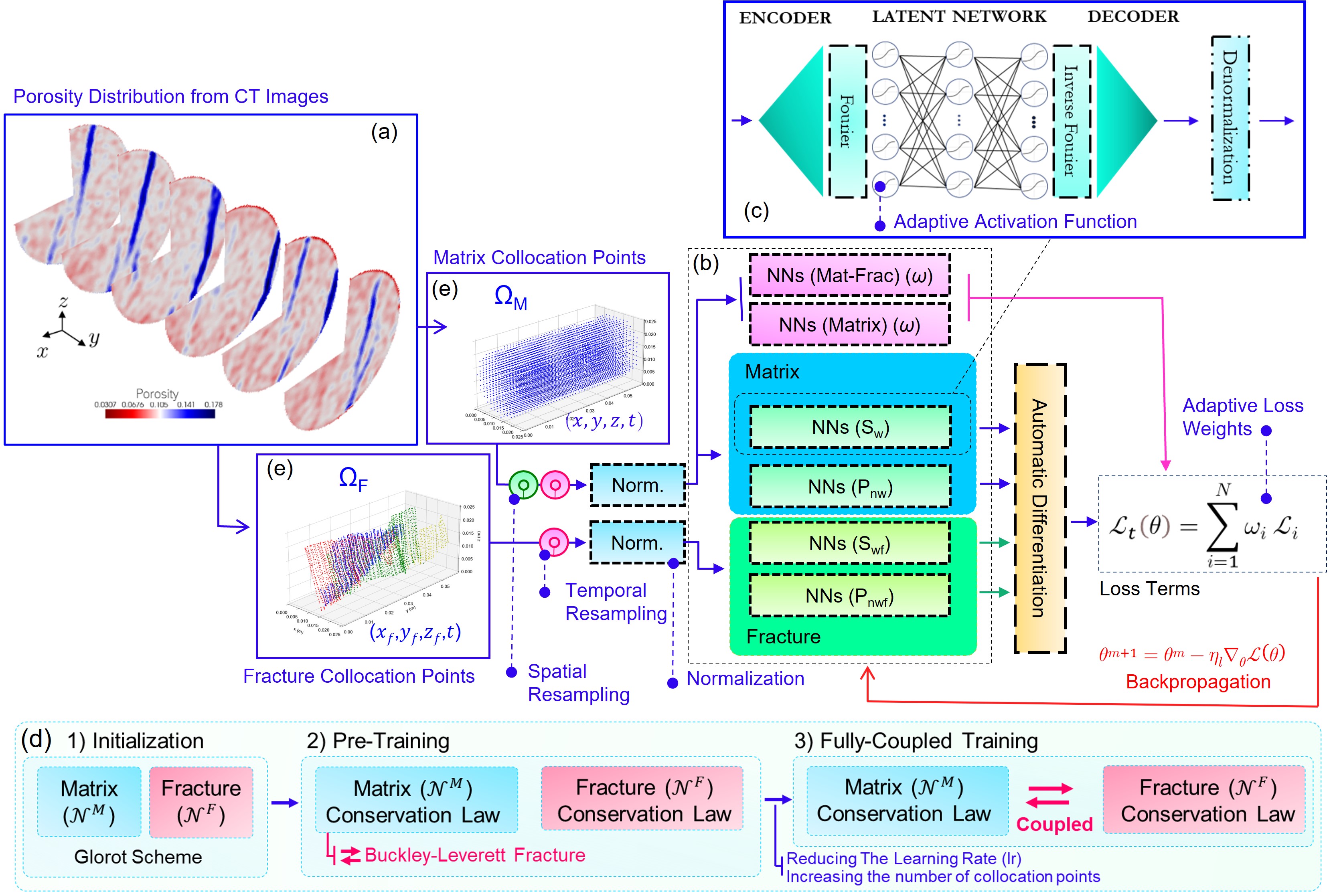}
\end{center}
\caption{\textbf{A schematic of the applied PINNs-based computational workflow.} \textbf{a)} the cross-sectional view of the porosity distribution in the experimental data, \textbf{b)} The model architecture; separate networks for the state variables of matrix and fracture systems, \textbf{c)} The architecture of each NNs, \textbf{d)} The proposed step-wise learning strategy,  \textbf{e)} The spatial distribution of the matrix and fracture collocation points.}
\label{fig:2}
\end{figure}

The architecture of the applied PINNs model is shown in Fig. \ref{fig:2}.
Due to the geometric complexities and non-linear flow profiles, we treated the matrix and fracture systems as distinct domains, each interacting through fluid exchange. Then, separate networks were allocated to each domain, to increase the flexibility of the networks in capturing the high-frequency trends in the solution of equations. Also, for each domain, the system state variables, i.e., pressure (\(p_{nw}\))  and water saturation (\(s_w\)) had separate networks (Fig. \ref{fig:2}b). Before passing the data to the network, the input data was normalised to the approximate range of (-1,1). The normalization/denormalization technique is discussed in Appendix \ref{sec:PINNsComputationalstrat}. 

In the NNs corresponding to the \(s_w\) values (for both matrix and fracture), we used the following sequence of operators: an encoder, a Fourier transformer, a latent multilayer perceptron (MLP) network, an inverse Fourier transformer, and a decoder network (as shown in Fig. \ref{fig:2}c). The outlet of the NNs is then denormalized to the range of actual values. The Fourier transformers were used to make the networks able to capture the high-frequency or multiscale behavior \citep{Wang2021OnNetworks} in the frontal regions of the saturation profiles. The NNs for the calculation of \(p_{nw}\) were similar, except that the Fourier and inverse Fourier transformations were not applied. The MLP network, which had a depth of five layers and a width of 80 neurons, was activated by an adaptive \(tanh\) activation function \cite{Jagtap2019AdaptiveNetworks}. 
The model was initialized using the Glorot initialization scheme \citep{Glorot2010UnderstandingNetworks}. Separate MLP networks were used to address the self-adaptive local weighting of the errors in PDE residuals for matrix and fracture domains. See Appendices \ref{sec:PINNsimplementation} and \ref{sec:PINNsComputationalstrat} for more details.

\subparagraph{Training:} 
The network was trained utilizing an Adam optimizer employing a full-batch approach and incorporating a gradually decreasing learning rate, from 0.0003, to 0.0001. We have applied a weight decay value of 0.0001. The number of gradient descent training steps (epochs) depended on the problem's complexity. Typically, 15,000 to 25,000 epochs were needed for stable error values. 

\begin{table}[ht]

\caption{The network properties for the synthetic benchmark model}
\begin{center}
    \label{tab:benchPINNsprops}
    \small 
    \begin{tabular}[t]{llll}
    \hline
    Property & Value & Property & Value  \\ \hline
    \(\mathcal{N}^M \) Width  &  80 & \(\mathcal{N}^M \) Depth & 8 \\ \hline
    \(\mathcal{N}^F \) Width  &  60 & \(\mathcal{N}^F \) Depth & 6 \\ \hline
    Activation Function  &  Adaptive \(tanh\) & Optimizer & Adam \\ \hline
    Learning Rate (lr)  & 2e-4 & Weight Decay & 1e-4 \\ \hline       
    Batch size 	& 36000 & Fourier Transform & Active for \(s_w\) \\ \hline   
    \end{tabular}
\end{center}
\end{table}

\subsection{The Solution Approach} 
\label{sec:SolutionMethod}
The investigated system of equations governing the matrix and fracture domains exhibits a complex interdependency. Training the model using a fully-coupled approach generates chaotic computational behavior as the interaction terms between the matrix and fracture domains are highly sensitive to the accuracy of the calculations performed in each domain individually. On the other hand, the invasion of fluids from the fracture into the matrix domain is a strong function of the interaction terms. Crucially, the differing loss landscapes of the matrix and fracture PINNs models result in distinct convergence behaviors. To mitigate these issues, alternative preconditioning techniques are necessary \citep{Krishnapriyan2021CharacterizingNetworks}. In this work, we suggested to apply a sequence of pre-training training steps before performing the fully-coupled training and then simulations. A schematic of the approach is visualized in Fig. \ref{fig:2}d: At first, the matrix and fracture networks are trained individually. Then, we shift to the fully-coupled training technique.

\subparagraph{Pre-training:} This stage leverages the model reduction technique \citep{Chung2017CouplingApproaches} as a means to tackle multiscale problems. To do so, we applied an independent training strategy for matrix and fracture at the beginning of training, before shifting to the fully-coupled training strategy. In this approach, instead of defining the matrix-fracture transfer to modify the PDEs of flow, we explicitly applied their impacts as the boundary conditions in the matrix as follows. 

\begin{itemize}
\item Matrix: The multiphase flow in the matrix is analyzed by assuming that the \(s_w\)  and the \(p_{nw}\) in the matrix at points of intersection with fractures are equal to the corresponding values in the fractures
\begin{align}
s_w = s_{w}^{F,BL},\ \ \ \ {\in\Omega}_{MF} \label{eq:pretrsw}
\end{align}
\begin{align}
p_{nw} = p_{nw}^{F,BL},\ \ \ \ {\in\Omega}_{MF} \label{eq:pretrpnw}
\end{align}
\(s_{w}^{F,BL}\) is calculated using the one-dimensional (1D) Buckley-Leverett (BL) theory, and \(p_{nw}^{F,BL}\)  is calculated using linear pressure distribution assumption. The Buckley-Leverett theory offers a fundamental analytical approach for forecasting the saturation profile and front velocity of the displacing fluid in two-phase flow in porous media. The BL theory is explained in Appendix  \ref{sec:BLequation}. 
\item Fracture: The PINNs model corresponding to the PDE of multiphase flow in fractures is solved by ignoring the interactions with matrix. 
\end{itemize}

The pre-training technique  considers many of the flow mechanisms, but neglects the viscous flow interactions between the matrix and fracture. The difference gets more critical when the permeability and capillary variation in matrix and fracture media is less significant. The mentioned mechanisms are naturally considered in the fully coupled approach. So, after pre-training of the networks, and reaching an equilibrium in the training process, we switch to the fully coupled technique. The mathematics behind coupling the matrix and fracture is previously explained in section \ref{sec:mathematicsofphysics}. 

In the case of inverse calculations, before starting the optimization of inverse variables, it is essential to partially train the networks so that they can learn the main characteristics of the problem. This approach has been successfully applied in previous studies, such as those by \citet{Zhang2022AnalysesNetworks} and \citet{Abbasi2024ApplicationTests}. At this step, the trainable inverse parameters are kept frozen for some limited epochs.

\subsection{Collocation Points}
\begin{figure}[h]
\begin{center}
\includegraphics[width=.99\textwidth]%
    {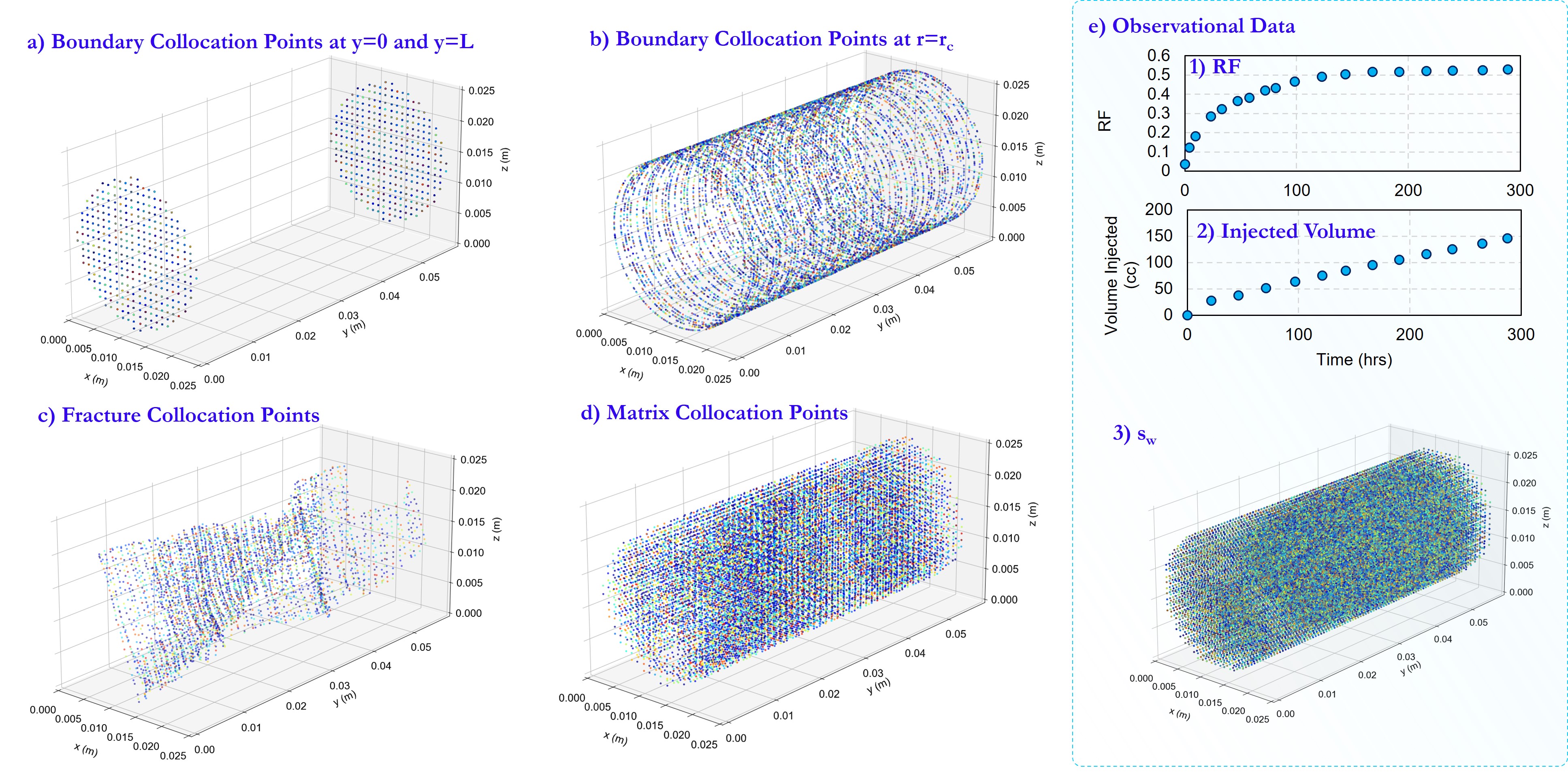}
\end{center}
\caption{ \textbf{The collocation points used for solving the problem under study.} The color of the points demonstrate the corresponding time for each point in the range \(t=1-10^6 \) sec. \textbf{a)} The collocation points at the boundary conditions of inlet (y=0) and outlet (y=L) faces (940 spatial points), \textbf{b)} The collocation points corresponding to the non-flowing boundary condition at \(r=r_{c}\) (8630 spatial points), \textbf{c)} The collocation points corresponding to the fracture (4103 spatial points), \textbf{d)} The collocation points corresponding to the matrix (23500 spatial points). The matrix collocation points with distance less than 0.0006 m from the fracture collocation points have been removed. \textbf{e)} the observational data points used in the data loss: 1) RF points, 2) points with the measured injected volume, and 3) spatiotemporal points of the in-situ \( s_w\) measured using CT-scan. }
\label{fig:colls}
\end{figure}

As it is visualized in Fig. \ref{fig:colls}, collocation points were extracted separately for the matrix and fracture domains based on the CT/micro-CT scan coordinates (also see Fig. \ref{fig:2}e). For the matrix domain, 23500 spatial collocation points were used in the core cylinder domain at specified x, y, z resolutions ($N$), excluding points in superposition with fractures. Boundary collocation points were collected for the cylinder ends and radius. Fracture collocation points were manually extracted from micro-CT scans ($N^F$). Temporal collocation points were randomly selected based on the \(\sqrt{t}\) distribution intervals, as it is expected that the system is mainly controlled by the spontaneous imbibition mechanism. \cite{Abbasi2023SimulationNetworks} have shown how the selection of temporal points based on \(\sqrt{t}\) distribution intervals helps in improving PINNs solutions. More detailed information about the experimental data and the collocation points are provided in Appendix \ref{sec:PINNsCollocation Points}.

\section{Results}

\subsection{Synthetic Benchmark Problem}
\label{sec:benchmarksection}
\begin{figure}[htbp]
\begin{center}
\includegraphics[width=.85\textwidth]%
    {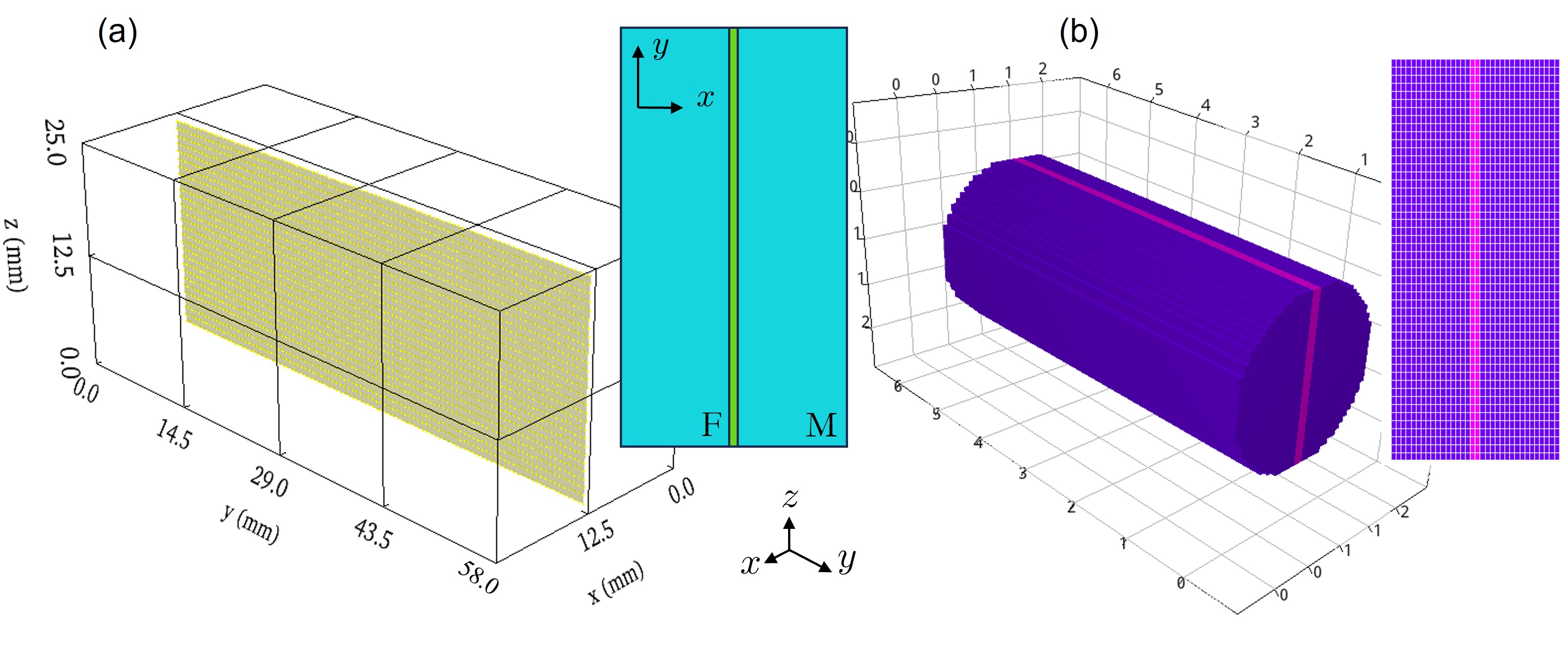}
\end{center}
\caption{\textbf{The geometry of the benchmark model,} \textbf{a)} The map of the fracture in the synthetic matrix-fracture problem, \textbf{b)} the equivalent static model for the FD simulations. }
\label{fig:BenchFracMap}
\end{figure}

This section focuses on validating the proposed PINNs-based workflow (section \ref{sec:SolutionMethod}) against forward and inverse modeling of a fully-characterized synthetic problem, corresponding to brine injection in a CO$_2$ saturated fractured porous media. Fig. \ref{fig:BenchFracMap}a provides a 3D visualization of a cylindrical core featuring a fracture (located at \(x=12.5\) mm) in the flow direction (\(y\)). The properties of the matrix and fracture are detailed in Table \ref{tab:benchprops}. Additionally, Fig. \ref{fig:BenchFracMap}b illustrates the static geometry of the matrix and fracture system, along with a 3D visualization of the discretized model used for the FD simulations. The applied simulator IORCoreSim by \citet{Lohne2017ARegimes} provided the true solution of the benchmark problem. The specifications of the utilized PINNs model is provided in Table \ref{tab:benchPINNsprops}. In this section, the network architecture is the same as in Fig. \ref{fig:2}. The pre-training strategy was applied in both forward and inverse problems.

\begin{table}[ht]

\caption{The properties of the synthetic benchmark matrix-fracture model. Wetting phase refers to water, and non-wetting phase refers to CO$_2$ phase.}
\begin{center}
    \label{tab:benchprops}
    \small 
    \begin{tabular}[t]{llll}
    \hline
    Property & Value & Property & Value  \\ \hline
    Matrix Porosity (-)  & 0.10 & \(k_{rw}^{\max}\)   &  0.20 \\ \hline       
    Matrix Permeability (mD)  &  0.000199 & \(k_{rnw}^{\max}\) & 0.20 \\ \hline
    Fracture Porosity (-) & 0.10 & \(n_{w1}\),\(n_{w2}\)    &  1.5, 1.5 \\ \hline
    Fracture Permeability (mD) & 0.0199 & \(n_{nw1}\),\(n_{nw2}\) & 2.0, 2.0 \\ \hline
    \(\mu_w\), \(\mu_{nw}\) (cP)  &  0.89, 0.0157 & IFT (N/m) & 0.04 \\ \hline
    \(J_1\) , \(J_2\)  &  0.02, 0.01 & \(s_{wc}\) , \(s_{nwr}\) & 0.0, 0.33 \\ \hline

    \end{tabular}
\end{center}
\end{table}

\begin{figure}[htbp]
\begin{center}
\includegraphics[width=.95\textwidth]%
    {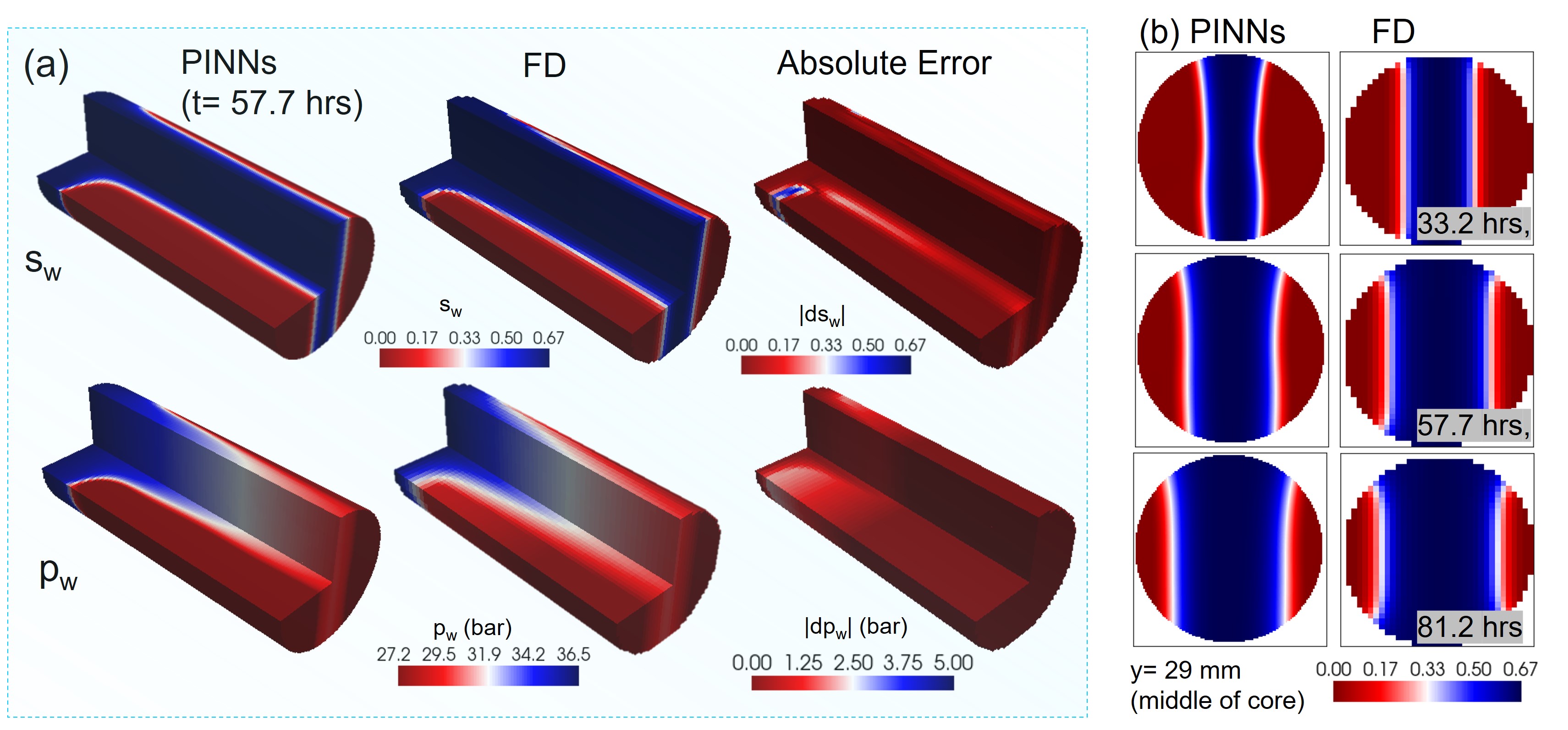}
\end{center}
\caption{\textbf{The comparative analysis of the frontal behavior of fluids as simulated by PINNs and FD for the benchmark problem forward simulation.}  \textbf{a)} a 3D visualization of  \(s_w\), and \(p_w\) profiles, and their point-wise absolute errors at t=57.7 hrs. \textbf{b)} the 2D cross-sectional view of \(s_w\) at y=29 mm for various time points: t=33.2, 57.7, and 81.2 hrs.}
\label{fig:benchprof}
\end{figure}

\subparagraph*{Forward simulation. } The flow saturation profile within the matrix, after the forward solution of the problem based on known parameter values, is visualised in Fig.~\ref{fig:benchprof}. The visualisation of the 3D profile of \(s_w\) and \(p_{w}\) as well as cross-sectional lateral invasion of water demonstrate the ability of the PINNs model to capture key 3D flow characteristics. The small errors in the predictions - i.e., the mean absolute error (MAE) of 0.028 for \(s_w\) and 0.42 bar for \(p_w\) - is accumulated mainly around the imbibition fronts. The computational times for the numerical solver and the PINNs solver were comparable for this specific forward problem. Applying the pre-training strategy was crucial to achieving high accuracy solutions, as demonstrated in Appendix \ref{sec:sensitivityanalysis}.

Fig.~\ref{fig:benchprof} clearly shows a sharp change in saturation at the flow fronts. The favorable mobility contrast, primarily due to the low viscosity of CO$_2$ compared to water (viscosity ratio of \(\mu_{\text{CO}_2}/\mu_{\text{water}} = 0.017\)) gives steep saturation profiles both during counter-current imbibition \cite{Andersen2023Early-Coefficient} and forced imbibition \cite{buckley1942mechanism}. Viscous flow creates shock fronts that are challenging to solve by many mathematical methods, including PINNs \cite{Fuks2020LIMITATIONSMEDIA}. In the matrix, the frontal shock is mainly cancelled out by capillary spreading effects. However, in the fracture domain, flow is primarily viscous. We assumed an insignificant value of capillary pressure in the fracture domain to mitigate the issue, without compromising accuracy or speed of computations.

\subparagraph{Inverse calculations based on history matching of recovery factor.}
Here, we utilise PINNs for the purpose of obtaining the flow parameters of the system, i.e., saturation functions \(k_r\), \(p_c\) of the matrix, by history matching of the observational data. Based on these curves we also calculate the resulting curve \(\Lambda (s_w)\) which refers to a normalized Capillary Diffusion Coefficient (CDC) curve after constant parameters such as porosity and permeability have been scaled. \(\Lambda (s_w)\) expresses how the saturation functions affect capillary flow within porous media and is a function of the \(k_r\) and \(p_c\) curves \cite{Andersen2023Early-Coefficient}, 
 \begin{align}
\Lambda = \mu_m \lambda_{nw} f_w \left( -\frac{dJ}{dS_w} \right)
\label{eq:Lambda}
\end{align}
where, \(\mu_m\) is the geometric mean of the phase viscosities \((\sqrt{\mu_w \mu_{nw}}\)). Spontaneous imbibition in a given geometry can be uniquely characterized by the capillary diffusion coefficient (CDC) curve \cite{Andersen2023Early-andCoefficient, Andersen2024TheGeometries}. In a 1D scenario, multiple matched sets of \(k_r\) and \(p_c\) curves yielding the same $\Lambda$ led to the same observed fluid recovery \cite{Abbasi2024ApplicationTests}.  Our interest here is to see the possibilities of curve determination in geometrically and flow-dynamically complex systems regarding curve determination from history matching. 

At this stage, we consider only the recovery factor (RF) curve as the observation data - which represents the volumetric fraction of the CO$_2$ phase expelled from the porous media over time. In this benchmark example, the porosity and permeability of the matrix, as well as all the properties of the fracture (including locations, porosity, permeability, and relative permeability curves) are assumed to be known prior to calculations. 

\begin{figure}[htbp]
\begin{center}
\includegraphics[width=.95\textwidth]%
    {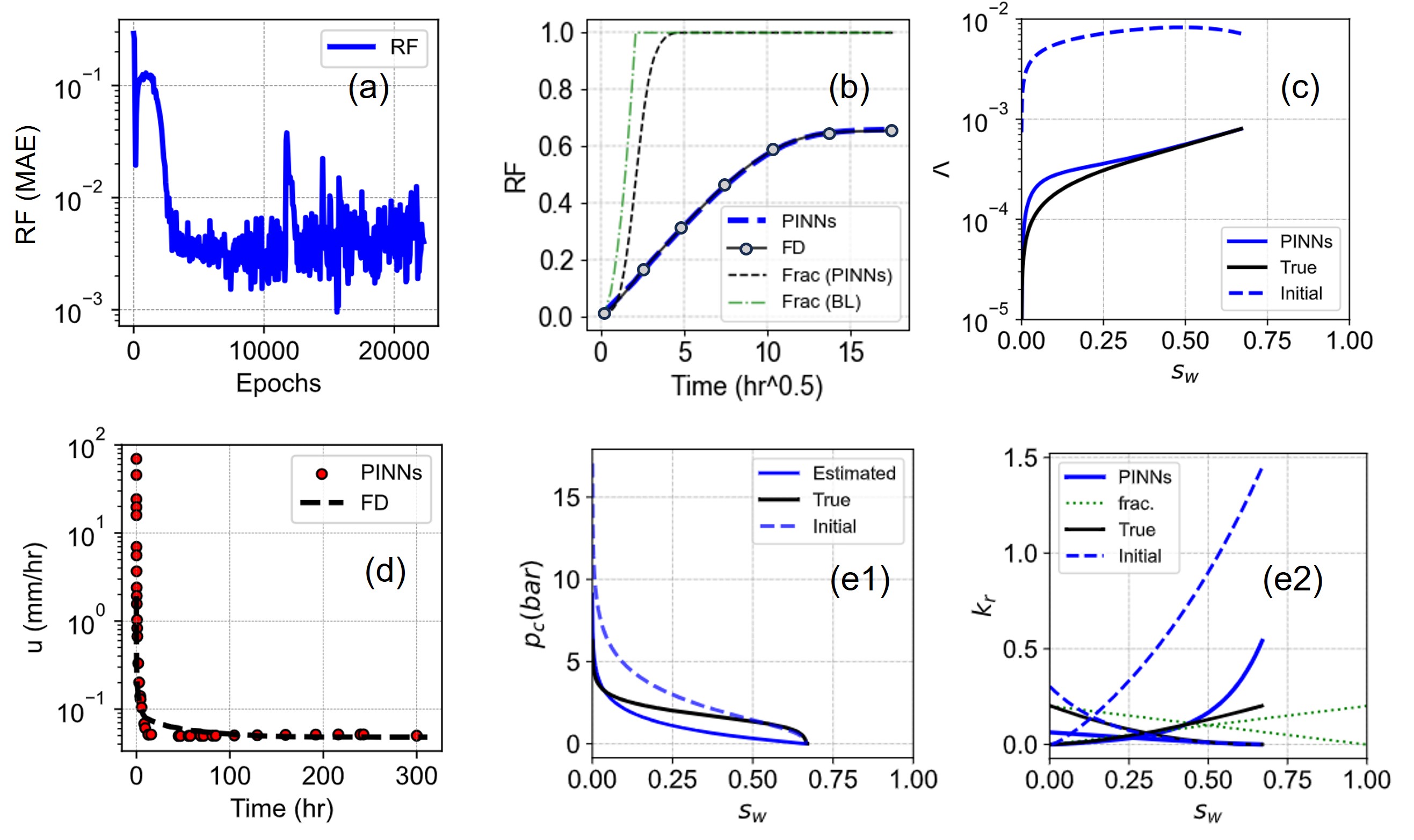}
\end{center}
\caption{\textbf{The results of inverse calculation for the synthetic benchmark model, based on history matching of the RF curve vs. time.} \textbf{a)} The trend of minimizing error compared to the RF data, \textbf{b)} The PINNs-based results of the RF curves for matrix and fracture, and compared to the true (FD) curve, \textbf{c)} The PINNs estimation of the \(\Lambda\) curve, compared to the randomly chosen initial guess and the true curve, \textbf{d)} The PINNs prediction for the total injection rate, compared to the true values.  \textbf{e)} The estimated flow curves compared to the true and the randomly initialized values: \( p_c\) and \(k_r\) curves, respectively. }
\label{fig:benchinvres}
\end{figure}

\begin{figure}[htbp]
\begin{center}
\includegraphics[width=.852\textwidth]%
    {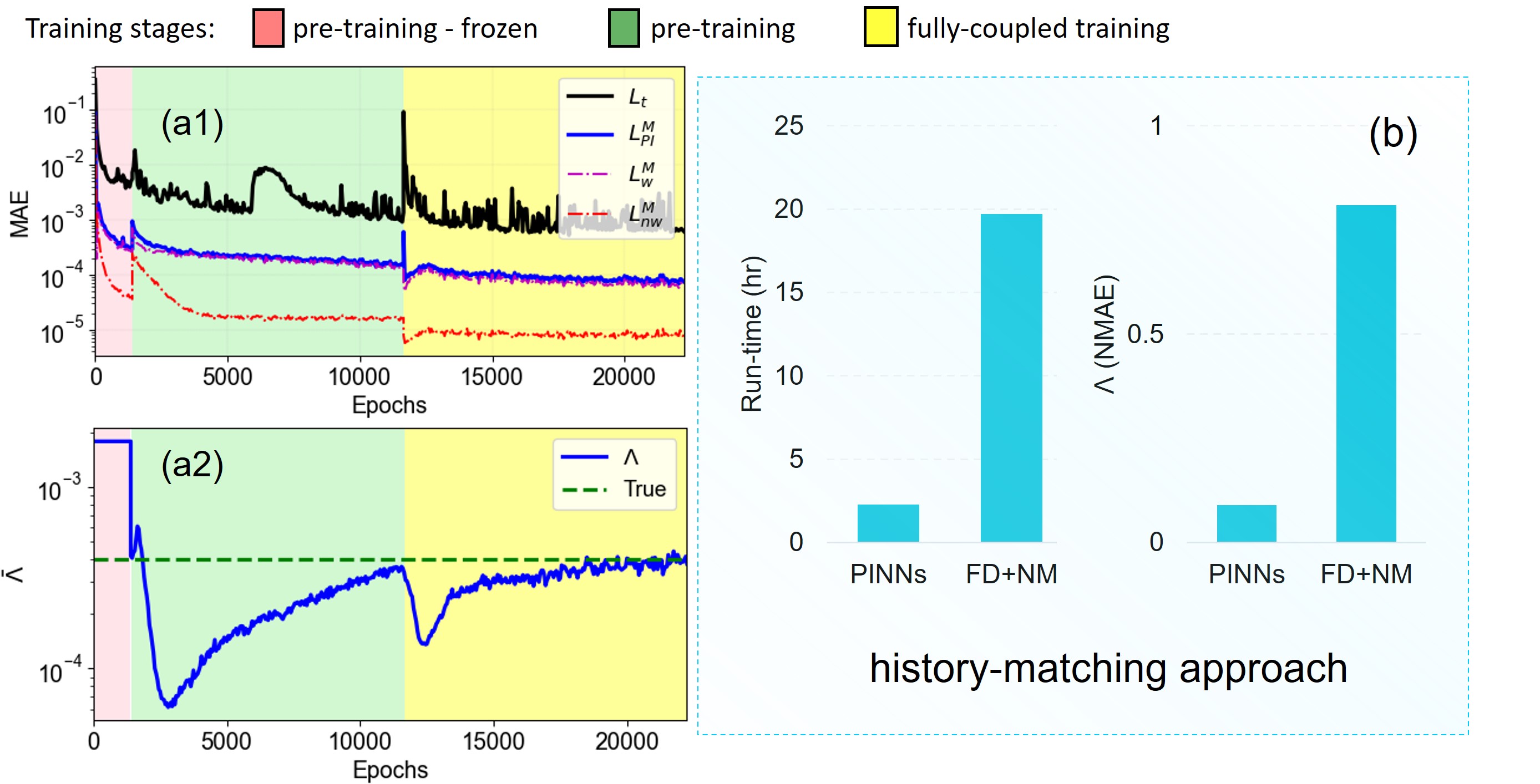}
\end{center}
\caption{ \textbf{The training of the PINNs model.}   \textbf{a)} 1) The total loss term, ${\mathcal{L}}_{t}$, and loss terms corresponding to the PDE residuals for matrix, i.e.,  ${\mathcal{L}}_{PI}^{M}$, ${\mathcal{L}}_{PIw}^{M}$, and ${\mathcal{L}}_{PInw}^{M}$ , 2) the trend of changes in $\bar{\Lambda}$, representing the area under $\Lambda$ curve, \textbf{b)} A comparison of the run-time and accuracy of two history-matching approaches in the estimation of the \(\Lambda\) curve: PINNs, and FD + Nelder-Mead (NM) optimizer.}
\label{fig:traininginverse}
\end{figure}

\begin{figure}[htbp]
\begin{center}
\includegraphics[width=.9\textwidth]%
    {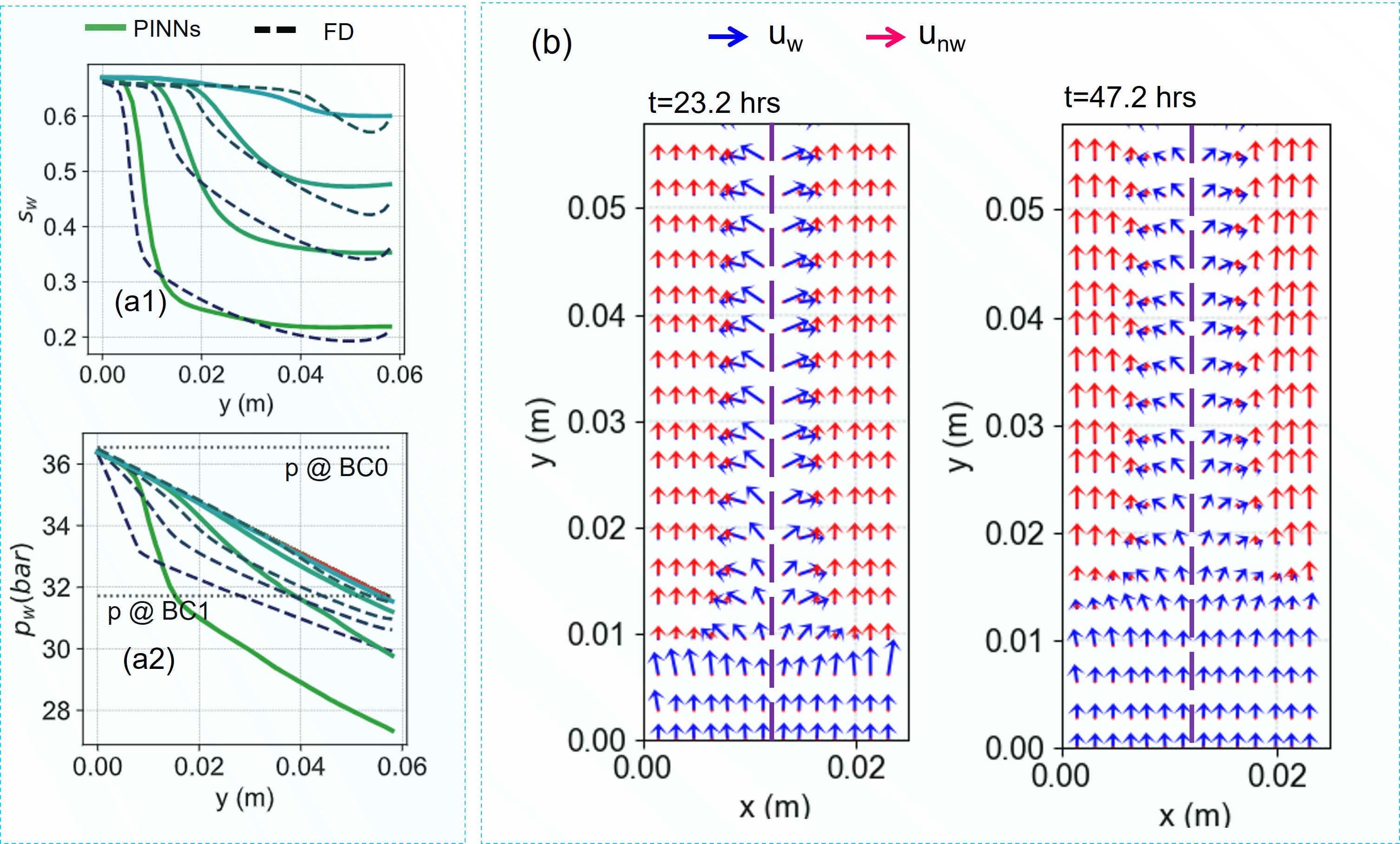}
\end{center}
\caption{ \textbf{ The profile of flow inside the matrix for the benchmark problem scenario, } \textbf{a)} The flow profile along the injection direction for (1) \(s_w\)  and (2) \(p_w\), at t= 23.22, 57.7, 98.2, 167.9 hrs,   \textbf{b)} 2D vector field plot at z=0.0125 m. Here, \( u_w \) and \( u_{nw} \) are not directly comparable, as they are scaled to the same range for visualization purposes.}
\label{fig:profileinverse}
\end{figure}

After random sampling of the inverse parameters, the calculations continued for 22000 epochs, until the error in RF was minimized (Fig. \ref{fig:benchinvres}a), the loss terms are stabilized (Fig. \ref{fig:traininginverse}a1), and the trend of changes in inverse parameters plateaued (Fig. \ref{fig:traininginverse}a2). The matching of the matrix RF curve (Fig. \ref{fig:benchinvres}b) could lead to the accurate estimation of the \(\Lambda\) curve with normalized mean absolute error (NMAE) of 0.098 (see Fig. \ref{fig:benchinvres}c). PINNs calculations could also predict accurately the volumetric rate of the injected water, as shown in Fig. \ref{fig:benchinvres}d. The main determining factor on the brine injected rate is $K_F$, as the fracture had orders of magnitude higher permeability compared to the matrix. 
The model did not accurately retrieve the \(k_r\) and \(p_c\) curves (Figs. \ref{fig:benchinvres}e1 and \ref{fig:benchinvres}e2) due to the system's high degrees of freedom and the limited information used for matching. This limitation is evident in Fig. \ref{fig:profileinverse}a, where the workflow could predict the saturation profile, but less accurately the pressure distributions. The solution accuracy is explored further in the next section. Fig. \ref{fig:profileinverse}b compares the vector field of velocity for both phases. Although the wetting phase continuously imbibes into the matrix, the non-wetting phase follows viscous forces within the matrix towards the outlet of the system, indicating \textit{the governance of the co-current spontaneous imbibition mechanism in the matrix in a large fraction of the injection period}.

\begin{figure}[h]
\begin{center}
\includegraphics[width=.999\textwidth]%
    {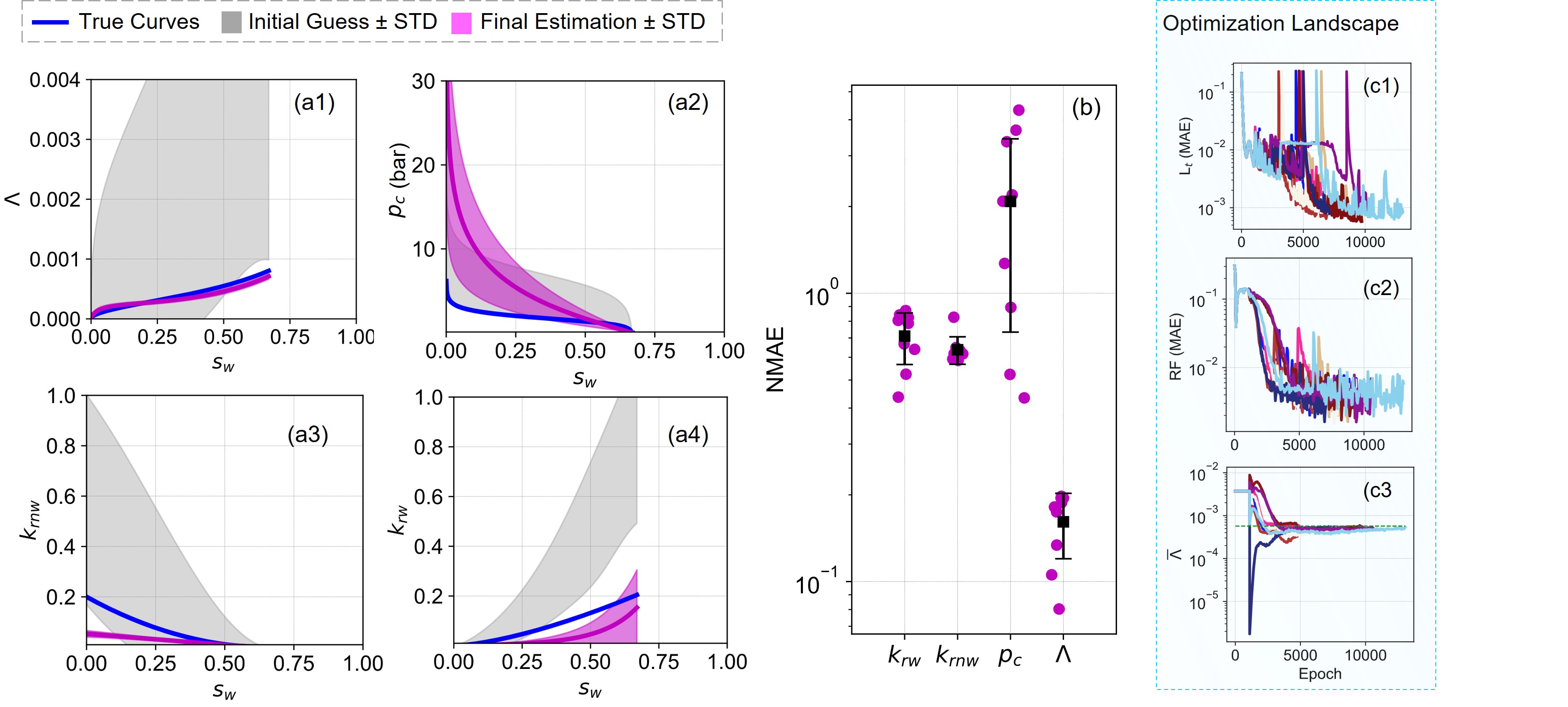}
\end{center}
\caption{ \textbf{The inverse calculation results for an ensemble with different initialization of the inverse parameters.} The initialization has been performed randomly.  \textbf{a)} the initial guess \textcolor{gray}{(gray)} and the final estimations after inverse calculations have been compared for \textcolor{magenta}{(magenta)} of \( \Lambda\), \( p_c\), \( k_{rw}\) and \( k_{rnw}\)  curves, respectively. The \textcolor{blue}{blue} curves show the true values. The shaded areas represent the mean $\pm$ standard deviations (STD).  \textbf{b)} Normalized MAE (NMAE) for different estimated parameters. Magenta dots represent individual data points, and black squares indicate the mean NMAE with error bars showing the STD. \textbf{c)} the optimization landscape of different terms during optimization during the training: \( \mathcal{L}_t\), RF (MAE), and \( \bar{\Lambda}\), respectively.}
\label{fig:benchinvens}
\end{figure}

\subparagraph{Uncertainty quantification.} We employed an ensemble-based uncertainty quantification approach, performing multiple history matches with random initializations of the inverse parameters (refer to Appendix \ref{sec:PINNsComputationalstrat} for detailed methodology). As demonstrated in Fig. \ref{fig:benchinvens}a and b, the \(\Lambda\) curves consistently converged, all with low errors compared to the true \(\Lambda\) curve. Fig. \ref{fig:benchinvens}c shows how the solution errors plateaued during training. Overall, the results shows that \textit{ the system is well defined  with respect to the \(\Lambda\) curve}, meaning that given the provided information there is only one \(\Lambda\) curve that can produce the observed measurements. In contrast, the \(k_r \) and \(p_c\) curves converged to distinct sets, exhibiting the high degree of freedom in the system of equations. Similar results have been discussed in \citet{Abbasi2024ApplicationTests} for a 1D spontaneous imbibition system. The results suggest that the significantly higher permeability of the fracture, compared to the matrix, allows it to become quickly saturated with water. As a result, the fracture-matrix contact surfaces behave like imbibition boundaries, leading to capillary-dominated flow between the matrix and fracture \cite{Berkowitz2002CharacterizingReview}.

\subparagraph{Computational cost.} The PINNs calculations were performed on a single A2000 GPU, which took around 1.5-2.5 hours depending on the size of the network and number of collocation points. The computational time was in the same range for both the forward and inverse problems, regardless of the fidelity of observational data used during history-matching. For comparison, we also attempted to history-match the problem using a coupled FD and Nelder-Mead (NM) optimizer (see Appendices \ref{sec:compres} and \ref{sec:historymatching} for more info). Individual forward simulations for a 27,000 cell FD model ranged from 40 minutes to 2 hours, depending on the complexity of effective parameters (e.g., flow parameter shapes and matrix-fracture property contrasts). As it is shown in Fig. \ref{fig:traininginverse}b, in the case of history matching the RF curve, the computational time of FD-NM optimisation was around 20 hours, with the NMAE of 0.81 (compared to the NMAE of 0.098 for the PINNs workflow) in the estimated $\Lambda$ curve. The results highlight \textit{the computational advantage of proposed PINNs-based workflow compared to FD-based history-matching approach for inverse evaluation of experiments of multiphase flow in fractured porous media}.

\subsection{Application of PINNs Workflow to a Real Experiment}
This section employs the PINNs workflow, with properties shown in Table \ref{tab:2}, for the interpretation and parameter identification of a real experiment of multiphase water-CO$_2$ flow in a naturally fractured shale rock sample, presented in section \ref{sec:experimentaldata}. The known properties of the system are shown in Table \ref{table:shalerockprops}. As an initial estimate, \(K\) and \(K_F\) are assumed to be 0.01 and 100 multipliers of the mean permeability of the rock measured experimentally (\( K = 0.01 K_{exp} \) and \( K_F = 100 K_{exp} \)). Without loss of generality, \( K \) is assumed constant during the inverse calculations, and the transmissibility of the matrix, as the most important unknown characteristics of the system, is adjusted based on the calculated \(\Lambda\) curve. The transmissibility of fracture is matched by altering \( K_F \) during the optimization, but assuming relative permeabilities and capillary pressure curves constant. In summary, among all properties of the system, \(k_r(s_w)\), \(p_c(s_w)\), and their combined alternative, \(\Lambda(s_w)\), as well as \(K_F\) are unknown and are required to be estimated based on the experimental observations.

\begin{table}[ht]

\caption{The network properties applied for the experimental problem. }
\begin{center}
    \label{tab:2}
    \small 
    \begin{tabular}[t]{llll}
    \hline
    Property & Value & Property & Value  \\ \hline
    \(\mathcal{N}^M \) Width  &  80 & \(\mathcal{N}^M \) Depth & 8 \\ \hline
    \(\mathcal{N}^F \) Width  &  60 & \(\mathcal{N}^F \) Depth & 6 \\ \hline
    Activation Function  &  Adaptive \textit{tanh} & Optimizer & Adam \\ \hline
    Learning Rate (lr)  & 0.10 & Weight Decay & 1e-4 \\ \hline       
    Batch size 	& 46000 & Fourier Transform & Active for \(s_w\) \\ \hline   
    Resampling Sequence 	& 10 & Temporal Sampling & Cartesian \\ \hline   
    \end{tabular}
\end{center}
\end{table}

\begin{table}[ht]

\caption{The properties of the experimental system corresponding to the core-flooding in the fractured shale rock. Wetting phase refers to water, and non-wetting phase refers to CO$_2$ phase. See Appendix \ref{sec:PINNsCollocation Points} for more information.}
\begin{center}
    \label{table:shalerockprops}
    \small 
    \begin{tabular}[t]{llll}
    \hline
    Property & Value & Property & Value  \\ \hline
    \( \phi\) (-)  & 0.10 & \( \phi_F\) (-) & 0.10 \\ \hline 
    \( l\) (cm)  & 5.8 & \( r_c\) (cm) & 1.25 \\ \hline       
    \(p_{in} \) (psi)  & 530 & \( p_{out}\) (psi) & 460 \\ \hline       
    
    $K$ (mD)  &  0.00019 & $K_F$ (mD) & 1.88 \\ \hline
    \(\mu_w\), \(\mu_{nw}\) (cP)  &  0.89, 0.0157 & IFT (N/m) & 0.04 \\ \hline
    \(\rho_w\), \(\rho_{nw}\) (\(kg/m^3\))  &  998.7, 78.9 & \( s_{wi}\) & 0.0 \\ \hline

    \end{tabular}
\end{center}
\end{table}

\subparagraph{History matching of the recovery curve and the total injection rate.} 
Inverse computations were started by randomly generated initial guess values for each parameter. To perform history matching of the RF curve and the injection rate data, we employed the step-wise calculation process, as introduced in section \ref{sec:SolutionMethod}. The training process was stopped after 17000 epochs, when the trainable parameters reached a plateau. The inverse-calculated solution parameters were validated by running a fine-grid FD numerical simulation based on these parameters, and comparing the obtained results. 

\begin{figure}[htbp]
\begin{center}
\includegraphics[width=.999\textwidth]%
    {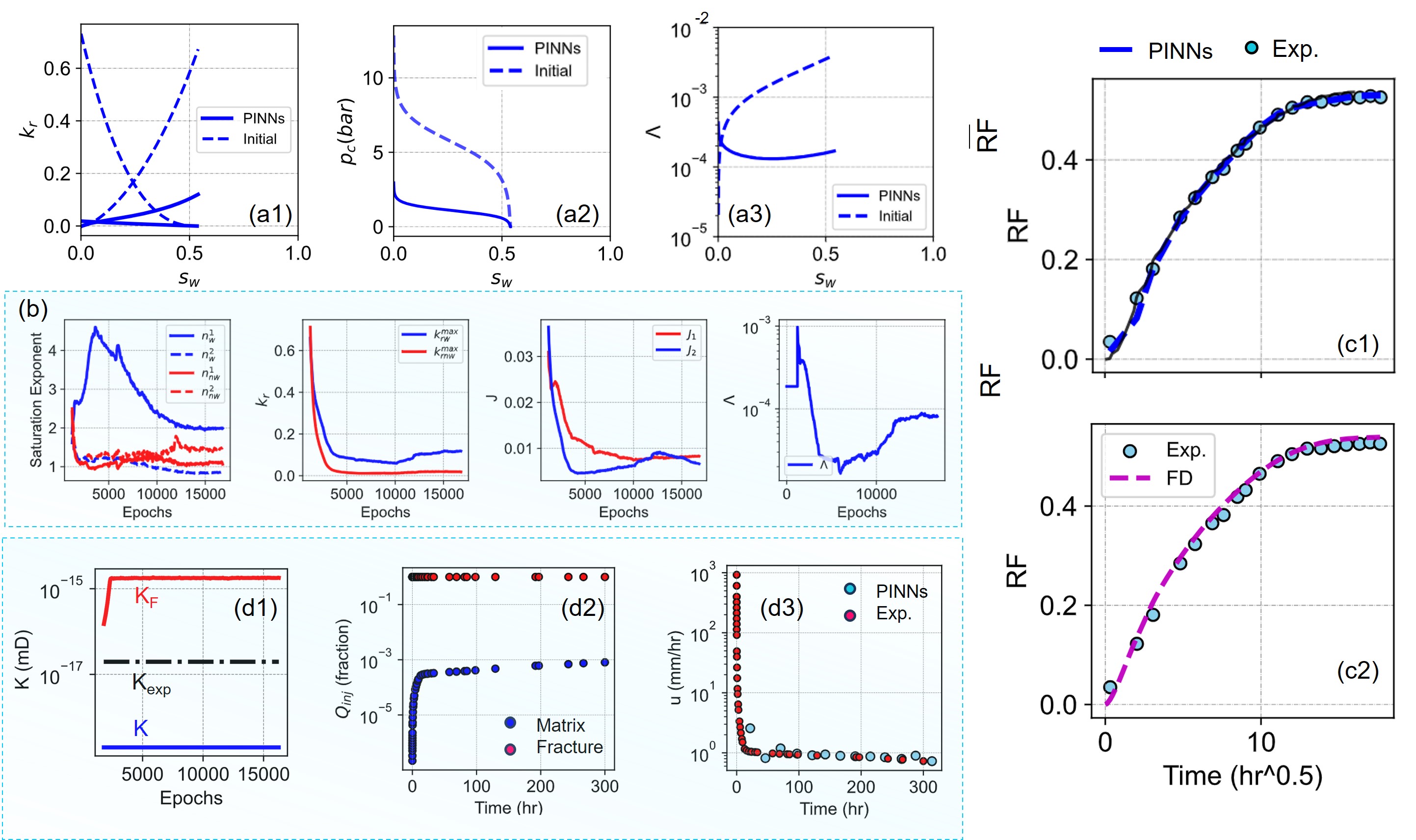}
\end{center}
\caption{ \textbf{The results for inverse calculation of matrix flow parameters by history matching of the RF experimental data.} \textbf{a)} The initial guess and final inverse calculated flow curves, i.e., \(k_r\), \(p_c\) and \( \Lambda\) curves, respectively. \textbf{b)} The trend of converging to the final values vs. epoch for different inverse parameters, \textbf{c)} The matched \( \bar{s_w}\) curve by PINNs (1) and the validation results from FD simulations (2), \textbf{d)} Matching the fracture permeability (\(K_f\)) and the history matching of volumetric injection rate: 1) the trend of changes in \(K_f\) versus the matrix permeability and experimentally measured permeability, \(K_{exp}\), 2) The contribution of matrix and fracture in the inlet face, 3) The total velocity of fluids compared to the experimental observational data.  }
\label{fig:expres1}
\end{figure}

\begin{figure}[htbp]
\begin{center}
\includegraphics[width=.99\textwidth]%
    {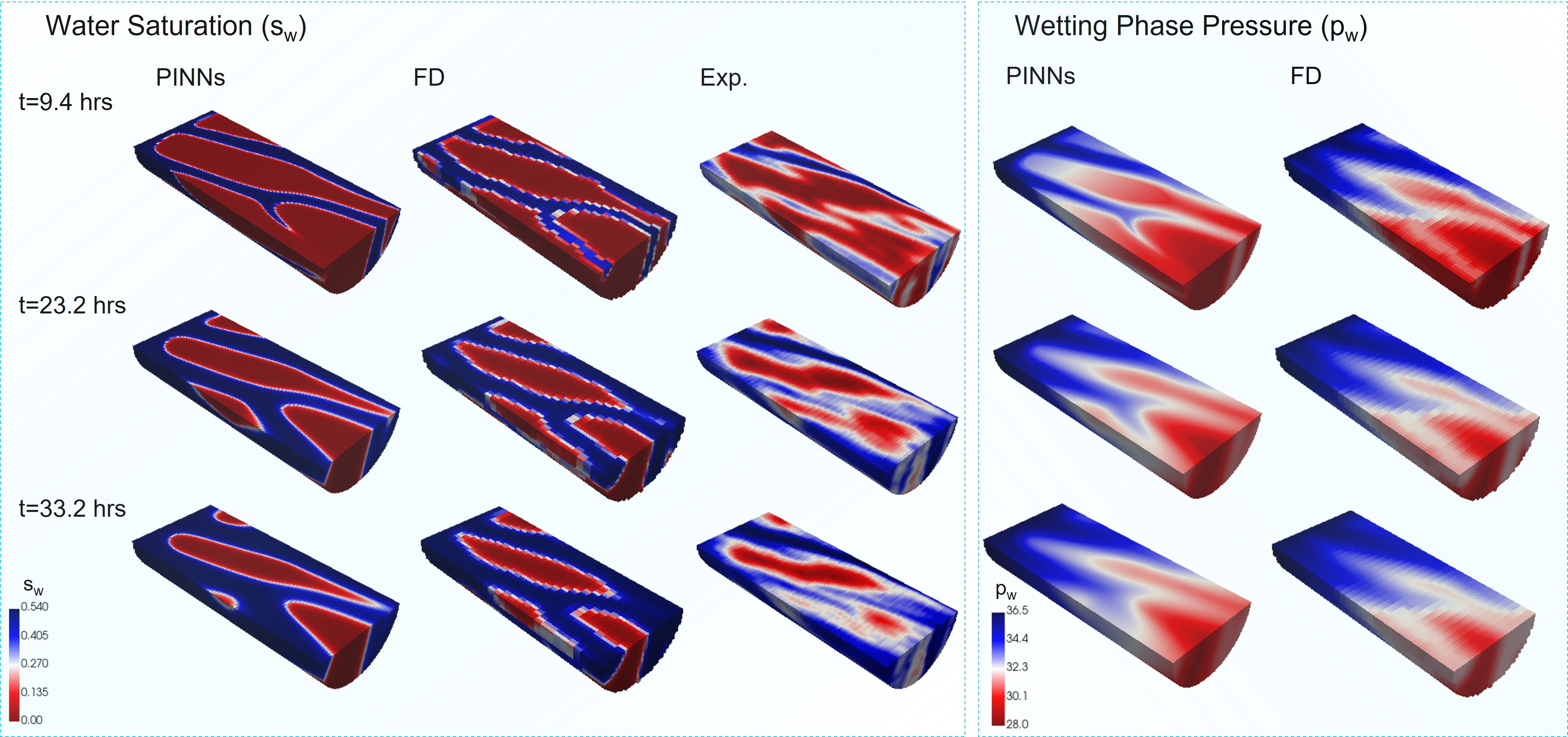}
\end{center}
\caption{ \textbf{A visualization compares water saturation ($s_w$) and wetting phase pressure ($p_w$) distributions in the porous media (matrix) at different times}. The figure compares the PINNs solution after inverse computations, FD solution after forward validation simulation, and experimental CT-scan images. It is important to note that the local saturation data was not utilized during the inverse calculations for the PINNs. Also, the FD simulation shows the forward analysis using the PINNs-based inverse calculated flow parameters. }
\label{fig:3DprofileExp}
\end{figure}

The inverse computed matrix flow parameters, i.e., \(k_r\), \(p_c\), (and resulting \(\Lambda\)) curves are shown in Fig. \ref{fig:expres1}a, after the trainable parameters reached a plateau (Fig. \ref{fig:expres1}b). The forward FD simulations with the obtained parameters show a reasonable match to the experimental RF data (Fig. \ref{fig:expres1}c). Fig. \ref{fig:3DprofileExp} compares the saturation and pressure profiles between PINNs and FD simulations. The profiles from the inverse calculations using PINNs closely match those from the FD simulations. Also, despite not using in-situ saturation data during training, the PINNs effectively captured the trends of water saturation in the CT images. Simultaneously, we matched the total injected fluids versus time, as shown in Fig. \ref{fig:expres1}d. The total volume of injected fluids is predominantly influenced by \(K_F\), as a significant portion $(>99\%)$ of the injected fluid enters the core through the fracture system and then flows toward the outlet face by bypassing the matrix. A small fraction of the fluid imbibes into the matrix (Fig. \ref{fig:expres1}d2), enabling the depletion of the non-wetting phase. The high injection rate at the beginning of the flow is due to the low flow resistance of the CO$_2$ phase (low viscosity), that initially saturated the fracture domain. The rate quickly drops to the level of $\approx$ 1.0 $mm/hr$ after breakthrough of the water phase at the outlet. A comparison of the RF curve, and the breakthrough time of water shows the injection rate of water is much higher than the critical rate of water advance; defined as the rate at which the positions of the water fronts in both the matrix and the fracture are equal \citep{Mattax1962ImbibitionReservoir}. It is expected that \textit{the frontal behavior of fluids in matrix is not significantly influenced by the total injection rate in the system}, due to the permeability dominance of the fracture domain \cite{Berkowitz2002CharacterizingReview}.

\begin{figure}
    \centering
    \includegraphics[width=0.75\linewidth]{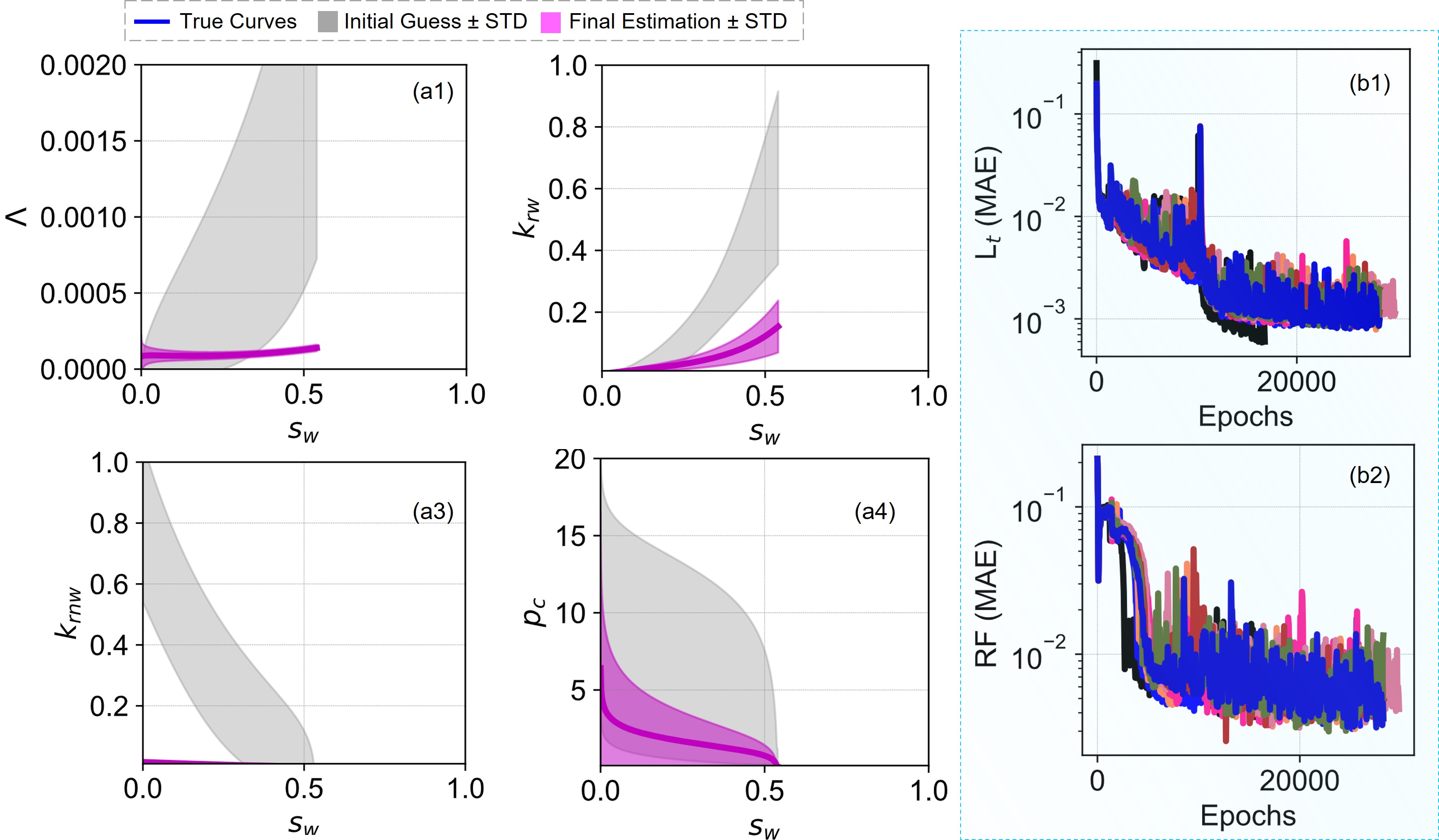}
    \caption{\textbf{The results of inverse calculations for an ensemble of models with different initialization values.}  a) the inverse calculated saturation functions. The grey curves are the initial guesses, and magenta curves exhibit the final estimations.  b) the convergence trends for total loss function, and the error in matched RF. }
    \label{fig:expensemble}
\end{figure}

\subparagraph{ The calculated $k_r$ and $p_c$ curves are not unique.} 
By performing ensemble-based uncertainty quantification - as the results are shown in Fig. \ref{fig:expensemble} - \textit{the \(k_r\) and \(p_c\) curves exhibit non-uniqueness}, while the \(\Lambda\) curve could be uniquely estimated, as expected based on the results of the previous sections. Uniquely estimating $k_r$ and $p_c$  would be more straightforward if viscous forces dominated matrix flow. However, in this system, matrix flow is primarily controlled by capillary forces, while flow in the fractures is governed by viscous forces, as outlined in the previous section.

\begin{figure}[htbp]
\begin{center}
\includegraphics[width=.8\textwidth]%
    {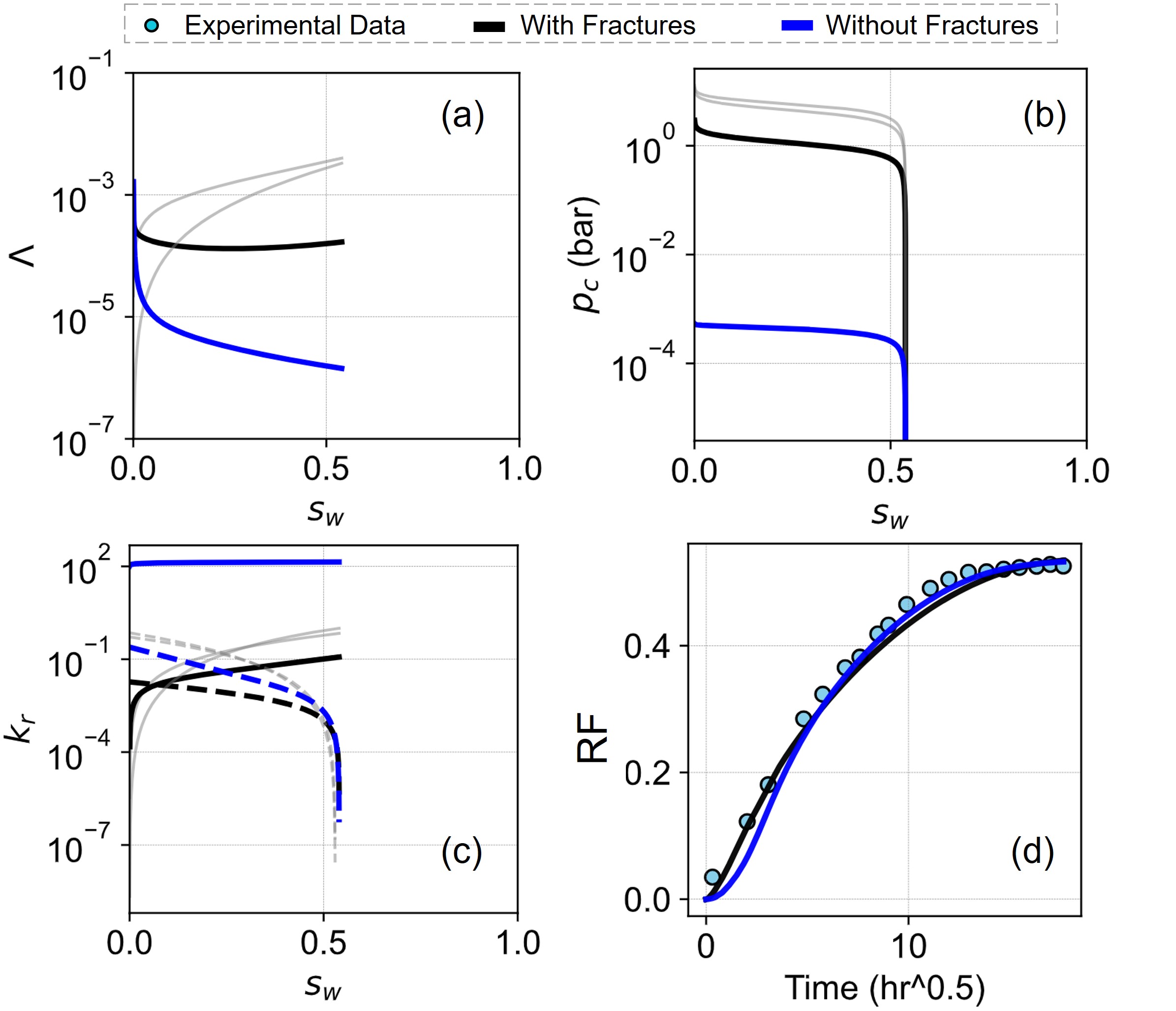}
\end{center}
\caption{ \textbf{The impact of multiscale simulations on the inverse-calculated flow parameters of porous media was examined in two cases: one that considered the impact of fractures and one that neglected it.} \textbf{a)} The inverse calculated \( \Lambda\) curve, \textbf{b)} The inverse calculated \( p_c\) curve, \textbf{c)} The inverse calculated \( k_r\) curve,  \textbf{d)} The validation of the obtained results for both cases with FD simulations, both match with the experimental RF data, }
\label{fig:fracimpact}
\end{figure}

\subparagraph{The importance of multiscale mechanisms.} 
We investigate the significance of modeling the multiscale mechanisms to understand multiphase flow in porous media under complex geometries, and their impacts on parameter identification. 
We performed inverse calculations for two separate cases — one that neglects fractures and their interactions with the matrix, and another that incorporates these fractures. In both cases, RF data and water injection rate (Q$_{inj}$) were used as observational data. As shown in Fig. \ref{fig:fracimpact}, a significant difference emerges when comparing the inverse-calculated saturation functions, even though both cases match the observational RF data (Fig. \ref{fig:fracimpact}d). 
This discrepancy led to unrealistic relative permeability and low capillary pressure values (as well as lower values in the $\Lambda$ curve), inconsistent with expected behavior in shale rocks, which normally exhibit high capillary and low permeability properties \cite{Rahmanian2010StorageReservoirs}. Consequently, viscous flow was incorrectly identified as the dominant mechanism, leading to a significant underestimation of capillary forces. These results underscore the critical importance of incorporating multiscale matrix-fracture interactions in flow simulations and using workflows capable of solving such problems efficiently. Current workflows often overlook these multiscale effects due to limited data quantifying their impact, as well as computational inefficiencies and stability issues in numerical methods.

\subparagraph{History matching of in-situ water saturation data.}

\begin{figure}[H]
\begin{center}
\includegraphics[width=.99\textwidth]%
    {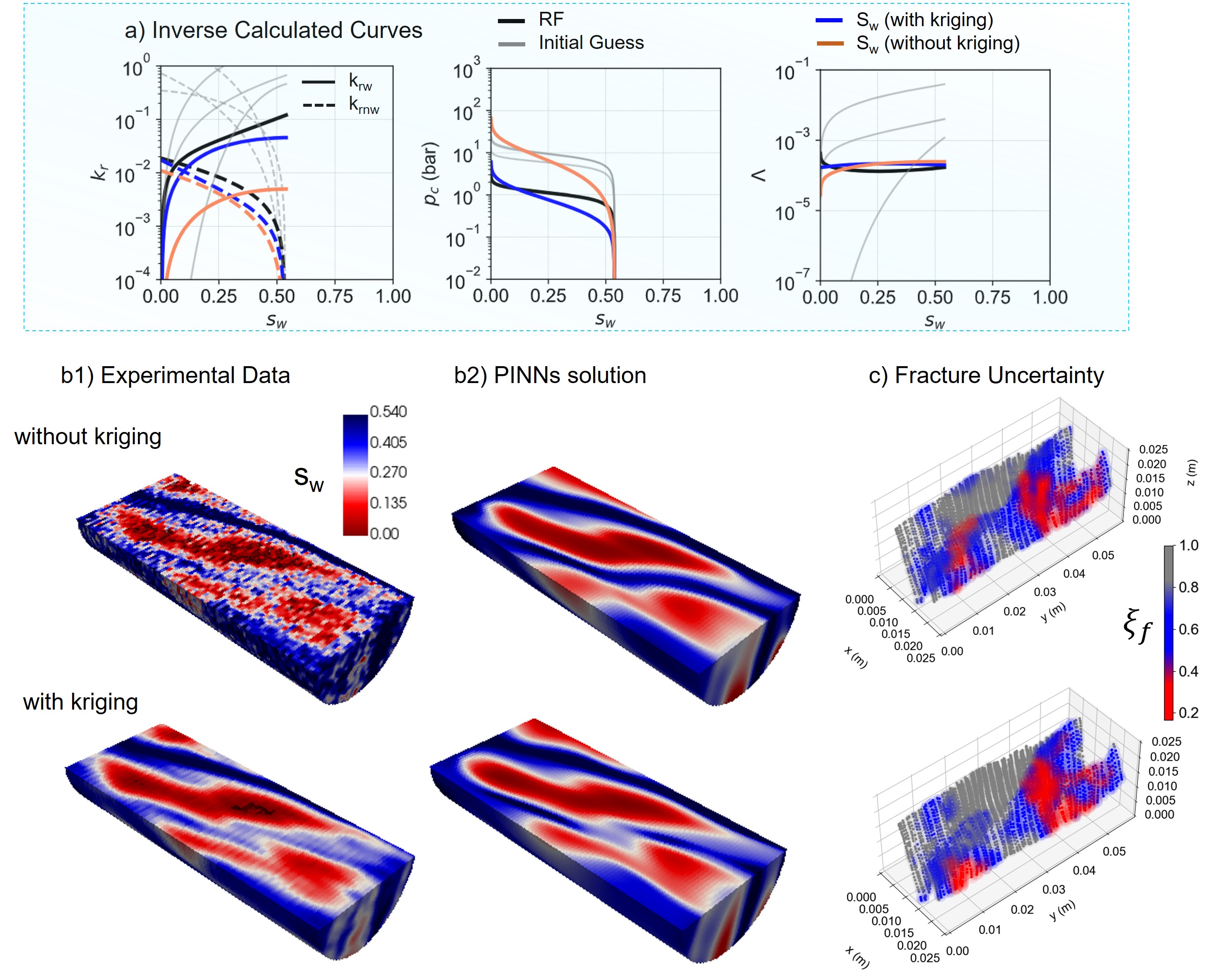}
\end{center}
\caption{\textbf{ The PINNs solution for inverse calculations when the in-situ water saturation data have been used as the observations.} A comparison has been made between two cases in which the observations have been pre-processed using a convolutional kriging algorithm. \textbf{a)} A comparison of the inverse calculated flow curves (\(k_r\), \( p_c\) and \(\Lambda\), respectively) for different scenarios of using observational data, \textbf{b)} A comparison of the 3D water saturation ($s_w$) profiles at $t=23.4$ hrs, \textbf{c)} The results of fracture uncertainty quantification; $\xi_f=1$ (\textcolor{gray}{gray}) indicates minimal uncertainty, $\xi_f=0.6$ (\textcolor{blue}{blue}) represents medium uncertainty, and $\xi_f=0.2$ (\textcolor{red}{red}) represents high uncertainty) }
\label{fig:insitusat}
\end{figure}

History matching of high-fidelity datasets, in this case (3+1)D in-situ water saturation data from CT scan images, presents a significant challenge but also offers exciting opportunities to gain deeper insights into porous media properties. In Fig. \ref{fig:insitusat}, rather than relying on RF data for the inverse calculations, we utilized the local spatiotemporal distributions of \(s_w\) directly as observational data for both history matching and the estimation of flow parameters. Additionally, we employed an adaptive strategy to assess fracture importance by detecting discrepancies between the calculated saturation distribution and high-fidelity CT-scan data. During the pre-training stage, we introduced a new loss term to minimize the error between the PINNs prediction of $s_w$ and the experimental saturation distribution. A trainable vector, $\xi_f$, was used as a self-adaptive multiplier to adjust local matrix-fracture interactions. Initialized at 1, $\xi_f$ balances fracture impacts, and its value is optimized to reduce discrepancies between predictions and CT-scan data. The approach is explained in Appendix \ref{sec:PINNsimplementation}.

A comparison of history matching of the datasets with and without denoising the observational data show no meaningful differences in both simulated saturation distributions or inverse calculated \(\Lambda\) curves. Denoising was performed via a 3D convolutional kriging approach, as explained in Appendix \ref{sec:PINNsComputationalstrat}. These results exhibit \textit{the capabilities of PINNs in extracting the governing physical properties of systems even in the presence of noisy observations} \citep{Chen2021Physics-informedData}. 

In Fig. \ref{fig:insitusat} c, the results of the fracture uncertainty assessment are shown. In the points with grey colour, the PDE solution closely matches the CT images, indicating high certainty in the defined fracture system. In the blue zones, the CT images show less imbibition than expected values by the PINNs solution, suggesting lower matrix-fracture interactions. Conversely, in the red regions, the saturation trends in the CT images differ significantly from the PINNs expectations, highlighting discrepancies that were successfully detected in both the original and denoised datasets. The results successfully showed the differences between the fracture properties, as well as uncertainties in the zones considered as the fractured areas. The results demonstrate that, while we assumed a connected fracture network with uniform properties, the observations may not confirm these assumptions.

\section{Conclusions}

In this study, we have proposed a computational workflow based on Physics-Informed Neural Networks (PINNs) as a solver for the simulation and parameter identification of (3+1) dimensional two-phase flow processes in naturally fractured porous materials. 
The workflow is quite general, and can be applied to interpret a wide range of experimental datasets involving multiphase flow in porous media with complex geometries. The PINNs model employed a multi-scale network architecture, along with pre-training strategies and various regularization techniques (e.g., adaptive weighting, collocation point resampling, and Fourier transforms) to solve the PDEs under investigation.
After validation of the model against a synthetic scenario, we applied the framework to interpret and analyse a multi-fidelity experimental dataset for water injection in a CO$_2$ saturated fractured shale rock. The studied experimental problem is characterized by a confluence of complexities: significant permeability contrasts, unpatterned distribution of fractures, highly unfavorable mobility ratios, and the presence of shock fronts in the fractures. The results of the study can be summarized as follows:

\begin{itemize}
    \item   The results demonstrated the effectiveness of the workflow in performing forward calculations by capturing the multiscale flow dynamics with mean absolute error (MAE) of 0.028 for water saturation, and  0.42 bar for the wetting phase pressure, mainly accumulated around the saturation front. 
    
    \item For the benchmark model, by including only the RF data, the inverse calculations achieved an accurate estimation of flow properties, with a normalized MAE of 0.098 for the matrix capillary diffusion coefficient curve ($\Lambda$), as well as the water injection rate. In the real experimental scenario, the matrix and fracture flow properties ($\Lambda$ and $K_F$) were estimated by matching both the RF data and the water injection rate. The estimated properties closely resembled the experimental data with acceptable accuracy after numerical validation.

    \item     The implemented uncertainty quantification analysis yielded valuable insights into the unique and non-unique characteristics of the obtained solutions. As the flow in matrix was mainly governed by capillary forces (imbibition), the problem could be uniquely characterized with a unique \(\Lambda\) curve, which  combines different flow properties of the system, confirming the relative dominance of capillary forces as the flow mechanism in the matrix. However, reaching reliable information on individual \(k_r\) or \(p_c\) curves was not possible due to the unfavorable degree of freedom of the system, as demonstrated by the applied uncertainty quantification analysis. Additionally, simultaneous history-matching of RF curve, and brine injection rate depicts that the fractures served as primary flow conduits in the system. So, we could reach unique values for fracture permeability. 

    \item   In the case of inverse calculations, the workflow performed the calculations order(s) of magnitude faster and more accurately than the finite-difference-based numerical simulation methods (coupled with a Nelder–Mead optimizer).     

    \item     Our comparison of inverse calculations with and without fractures demonstrates the pivotal role of multiscale phenomena in interpreting the experiments of imbibition in fractured porous media, as the presence of fractures introduces a hierarchy of scales in the system. These multiscale effects, while commonly overlooked in practical procedures, can significantly change the overall system behavior, influencing the inverse calculated parameters such as effective permeability, capillary pressure, and relative permeability relationships, as well as the interplay between different flow mechanisms in the system. For instance, the role of capillary forces may be significantly mischaracterized, as the harmony between capillary forces in the matrix and viscous forces in the fractures creates a complex, scale-dependent flow regime. This oversight can result in misleading parameter estimations, misattribution of flow behaviors, and ultimately, the development of incorrect conceptual or large-scale geological models, potentially leading to poor predictions and decision-making in practical applications. 
    
    \item     The proposed workflow, along with the applied strategies, performed effectively in evaluating the (3+1)D noisy high-fidelity water saturation observations, achieving results comparable to those from the denoised dataset. This results demonstrates the capabilities of PINNs in capturing the noisy observations. Additionally, discrepancies between the applied matrix-fracture collocation points and the CT-scan images were successfully detected, providing a method for quantifying uncertainties in the presumed matrix-fracture interactions.
\end{itemize}

At the end, while the workflow offers more accurate and order(s) of magnitude quicker solutions in comparison to a standard approach, further development is necessary to achieve fully reliable models for a wider range of multiphase flow scenarios. Increasingly complex fracture distributions likely require even more complex neural network architectures. This highlights the need for developing more flexible network architectures capable of handling the significant nonlinearities existing in such complex systems.


\section*{Acknowledgments}
Abbasi, Andersen, and Hiorth gratefully acknowledge the support of the Research Council of Norway and the industry partners of NCS2030 (RCN project number 331644). They also thank the Research Council for funding the "Pisces-AI: Physics-Informed AI for Subsurface Characterization Experiments" project (RCN project number 354776).

\section*{Nomenclature}

\begin{table}[H]
    \centering
    \begin{tabular}{ll}
        \hline
        \textbf{Symbol} & \textbf{Description} \\
        \hline
        $\Omega_M$ & Matrix domain \\
        $\Omega_F$ & Fracture domain \\        
        $K$ & Absolute matrix permeability (m$^2$) \\
        $K_F$ & Fracture permeability (m$^2$) \\
        $k_{ri}$ & Relative permeability of phase $i$ (dimensionless) \\
        $\lambda_i$ & Mobility of phase $i$ \\
        $\mu_i$ & Viscosity of phase $i$ (Pa·s) \\
        $p_i$ & Pressure of phase $i$ (Pa) \\
        $p_w$ & Wetting phase pressure (Pa) \\
        $p_{nw}$ & Non-wetting phase pressure (Pa) \\
        $p_c$ & Capillary pressure (Pa) \\
        $s_i$ & Saturation of phase $i$ \\
        $s_w$ & Water saturation (dimensionless) \\
        $s_{nw}$ & Non-wetting phase saturation (dimensionless) \\
        $s_{wc}$ & Connate water saturation (dimensionless) \\
        $s_{nwr}$ & Residual non-wetting phase saturation (dimensionless) \\
        $\phi$ & Porosity (dimensionless) \\
        $\rho_i$ & Density of phase $i$ (kg/m$^3$) \\
        $u_i$ & Darcy velocity of phase $i$ (m/s) \\
        $u_i^F$ & Fracture parallel velocity of phase $i$ (m/s) \\
        $v_i^\bot$ & Matrix-fracture velocity (m/s) \\
        $q_i^\bot$ & Mass transfer rate between matrix and fracture (kg/m$^3$/s) \\
        $V_i^\bot$ & Cumulative volume transfer between matrix and fracture (m) \\
        $Q_i^\bot$ & Cumulative Mass transfer between matrix and fracture (kg/m$^3$) \\
        $J(S_w)$ & Leverett J-function (dimensionless) \\
        $\Lambda$ & Capillary diffusion coefficient (dimensionless) \\
        $Bo$ & Bond number (dimensionless) \\
        $e_V$ & Fracture aperture (m) \\
        $e_h$ & Hydraulic fracture aperture (m) \\
        $P_{in}$ & Inlet pressure (Pa) \\
        $P_{out}$ & Outlet pressure (Pa) \\
        $L$ & Core length (m) \\
        $r_c$ & Core radius (m) \\
        $\sigma$ & Interfacial tension (N/m) \\
        $J_1, J_2$ & Leverett function parameters \\
        $n_w, n_{nw}$ & Saturation exponents for wetting and non-wetting phases \\
        $t$ & Time (s) \\
        \hline
    \end{tabular}
\end{table}



\newpage

\maketitle

\section*{Supplementary Materials}

\setcounter{section}{0}
\setcounter{figure}{0}  
\setcounter{table}{0}  
\pagenumbering{arabic}

\newpage

\section{Physics-Informed Neural Networks: Implementation }
\label{sec:PINNsimplementation}

In this section, we explore the details of the implementation of the PINNs model for modelling multiphase flow in porous media, including the specific formulations of each loss term. As the scales of pressure data and water saturation data differ significantly, we needed to ensure they are comparable by bringing them to similar scales. To achieve this, we normalized the pressure loss terms by dividing all related terms by \(\kappa_p\), defined as
\begin{align}
{\kappa}_{p} = ({p}_{in} + {p}_{out} ) /2
\label{eq:p2}
\end{align}

Furthermore, since the scale of loss terms related to PDE residuals also deviates significantly from the [0,1] range, we defined another PDE normalization multiplier as
\begin{align}
{\kappa}_{r} = \frac{ t_{max}}{({\rho}_{w} + {\rho}_{nw} ) /2} 
\label{eq:p2}
\end{align}
where \( t_{max}\) is the time related to the end of experiment, which is almost 1e6 sec (277.78 hrs).

\subsection{Flow in Matrix Domain}

The solution for the governing equations within the matrix domain is achieved by minimizing a weighted combination of loss terms, denoted as \(\mathcal{L}_{t}^{M} \). These loss terms account for the residuals of the PDEs, the initial conditions, and the boundary conditions applied at different boundaries.

\subparagraph{PDE residuals:} To solve the flow in the matrix, we must simultaneously address the conservation laws for both the wetting and non-wetting phases. Thus, we define different loss terms according to equation (\ref{eq:matrix_frac_cont})
\begin{align}
{{\mathcal{L}}}_{w}^{M} = \frac{1}{\kappa_r} \text{MAE}(\text{ }\phi{\partial }_{t}\left({\rho}_{w}{s}_{w}\right) + \nabla \left({\rho}_{w}{u}_{w}\right )\text{ },\text{ } 0.0),\text{ } \in {\Omega}_{M}
\label{eq:matrix_loss_w}
\end{align}
\begin{align}
{{\mathcal{L}}}_{nw}^{M} = \frac{1}{\kappa_r} \text{MAE}(\text{ }\phi{\partial }_{t}\left({\rho}_{nw}{s}_{nw}\right) + \nabla \left({\rho}_{w}{u}_{nw}\right)\text{ },\text{ } 0.0),\text{ } \in {\Omega}_{M}
\label{eq:matrix_loss_w}
\end{align}
Here, MAE is the Mean Absolute Error, a metric used to measure the errors. It represents the average absolute difference between the predicted values and the actual values:
\begin{align}
{\text{MAE}(\hat{y}_i, y_i) = \frac{1}{n} \sum_{i=1}^{n} \left| \hat{y}_i  - y_i \right|,}
\label{eq:maeeq}
\end{align}
where \( n\) is is the number of observations or collocation points, \(y_i\) is the actual value, and \( \hat{y}_i\) is the predicted value. The number of collocation points in matrix and fracture domains are $N$, and $N^F$, respectively. Then, the total loss corresponding to the residual of equations of flow in matrix is defined as
\begin{align}
{{\mathcal{L}}}_{PI}^{M} =  ({\mathcal{L}}_{w}^{M}  + {\mathcal{L}}_{nw}^{M}) \cdot e^{\omega^M}, \text{  } \text{  } \in {\Omega}_{M} 
\label{eq:volbalance}
\end{align}

where \( \omega^M\) represents the local weight modifier obtained by self-attention mechanism as outlined in Appendix \ref{sec:PINNsComputationalstrat}.  A loss term is defined to constraint the values of the self-adaptive modifier for
\begin{align}
{{\mathcal{L}}}_{\omega}^{M} = \text{MAE}(\omega^{M} , 0.0),\text{  } \in {\Omega}_{F}
\label{eq:»LomegaM}
\end{align}

\subparagraph{Matrix-fracture mass transfer:} For the matrix collocation points, where mass transfer occurs between the matrix and the superimposed fracture network, we formulate the residual term of the governing partial differential equation as follows

\begin{align}
{{\mathcal{L}}}_{w}^{MF} = \frac{1}{\kappa_r}\text{MAE}(\text{ }\phi{\partial }_{t}\left({\rho}_{w}{s}_{w}\right) + \nabla \left({\rho}_{w}{u}_{w}\right)  - ~\frac{{\rho}_{w}}{{e}_{V}}\left(2{v}_{w}^{\bot }\right)\text{ },\text{ }0.0),\text{  } \in {\Omega}_{MF}
\label{eq:matrix_loss_w}
\end{align}
\begin{align}
{{\mathcal{L}}}_{nw}^{MF} = \frac{1}{\kappa_r}\text{MAE}(\text{ }\phi{\partial }_{t}\left({\rho}_{nw}{s}_{nw}\right) + \nabla \left({\rho}_{w}{u}_{nw}\right)  - ~\frac{{\rho}_{nw}}{{e}_{V}}\left(2{v}_{nw}^{\bot }\right)\text{ },\text{ }0.0),\text{  } \in {\Omega}_{MF}
\label{eq:matrix_loss_w}
\end{align}

The interaction term is governed by the pressure differential between the matrix and fracture and the local transmissibility of the matrix. In the fully-coupled training stage, we then define

\begin{align}
{{\mathcal{L}}}_{PI}^{MF} =  ({\mathcal{L}}_{w}^{MF}  + {\mathcal{L}}_{nw}^{MF} ) \cdot e^{\omega^{MF}}, \text{  } \text{  } \in {\Omega}_{MF} 
\label{eq:LMFvolbalance}
\end{align}

Here, \( \omega^{MF}\) is defined as a local self-adaptive weight modifier. A loss term is defined to constraint the values of the self-adaptive modifier for
\begin{align}
{{\mathcal{L}}}_{\omega}^{MF} = \text{MAE}(\omega^{MF} , 0.0),\text{  } \in {\Omega}_{MF}
\label{eq:»LomegaMF}
\end{align}

\subparagraph{Pre-training stage:} 
In the pre-training stage, we assumed that the matrix collocation points corresponding to the fractured zones are acting as the boundary conditions of the system, following the pressure and saturation trends of the fractures. The pressure and saturation values versus time in fractures, however, were calculated via Buckley-Leverett technique. We then directly assigned the fracture saturation and pressure properties to the collocated matrix points,
\begin{align}
{{\mathcal{L}}}_{PT,s_w}^{MF} =  \text{MAE}(\text{ } ({s}_{w} - {s}_{w}^{F}) \xi_f \text{ },\text{ } 0.0 \text{ }), \text{  } \text{  } \in {\Omega}_{MF} 
\label{eq:LMFPTsw}
\end{align}
\begin{align}
{{\mathcal{L}}}_{PT,p_{nw}}^{MF} =  \frac{1}{\kappa_p}\text{MAE}(\text{ } ({p}_{nw} - {p}_{nw}^{F}) \xi_f\text{ },\text{ } 0.0  \text{ }), \text{  } \text{  } \in {\Omega}_{MF} 
\label{eq:LMFPTpnw}
\end{align}
where, \( {s}_{w}^{F} \) and \( {p}_{nw}^{F}  \) are the fracture state variables calculated via Buckley-Leverett and linear pressure drop equation, explained in Appendix \ref{sec:BLequation}. Also, \( \xi_f \) is defined as a self-adaptive multiplier acts as the fracture uncertainty identifier. Technically, it adjusts the local matrix-fracture interactions. \( \xi_f \) is a trainable vector defined for all fracture collocation points, and initialized to be equal to one. Then, a restricting loss terms is defined to balance the \( \xi_f \) impacts during the training
\begin{align}
{{\mathcal{L}}}_{\xi_f}^{MF} =  \text{MAE}(\text{ }  \xi_f \text{ },\text{ } 1.0   \text{ }), \text{  } \text{  } \in {\Omega}_{MF} 
\label{eq:LMFPTxi}
\end{align}

In trying to minimize the trade-offs between the loss terms comparing the PINNs predictions vs. CT-scan images, as well as the matrix-fracture physical constraints, the values of \( \xi_f \) is adjusted in a way that the lowest discrepancies was obtained. In scenarios where uncertainty quantification is not a focus, setting \( \xi_f = 1.0 \) for all points simplifies the optimization process.

\subparagraph{Initial conditions:}
With the initial state of the system fully defined, we can now express the corresponding loss terms as
\begin{align}
{{\mathcal{L}}}_{IC,s_w}^{M} = \text{MAE}({s}_{w}(x,y,z,t = 0) , {s}_{wc}) ,\text{  } \in {\Omega}_{M}
\label{eq:»LeqICmatrix}
\end{align}
\begin{align}
{{\mathcal{L}}}_{IC,p_{nw}}^{M} = \frac{1}{\kappa_p}\text{MAE}({p}_{nw}(x,y,z,t = 0) , {p}_{i}) ,\text{  } \in {\Omega}_{M}
\label{eq:»LeqICmatrix}
\end{align}

\subparagraph{Boundary conditions at inlet:}
At the inlet, the core is exposed to water under a constant injection pressure, allowing us to express the loss terms as
\begin{align}
{{\mathcal{L}}}_{BC0,s_w}^{M} = \text{MAE}({s}_{w}(x,y = 0,z,t),{s}_{w,max}),\text{  } \in {\Omega}_{M}
\label{eq:LBC}
\end{align}
\begin{align}
{{\mathcal{L}}}_{BC0,p_{nw}}^{M} = \frac{1}{\kappa_p}\text{MAE}({p}_{nw}(x,y = 0,z,t),{p}_{in}),\text{  } \in {\Omega}_{M}
\label{eq:LBC}
\end{align}

\subparagraph{Boundary conditions at outlet:}
To ensure proper flow characteristics, the outlet should be modeled with a constant pressure boundary condition
\begin{align}
{{\mathcal{L}}}_{BC1,pnw}^{M} = \frac{1}{\kappa_p}\text{MAE}({p}_{nw}(x,y = L,z,t),{p}_{out}), \text{  } \in {\Omega}_{M}
\label{eq:LpnwBC1}
\end{align}

\subparagraph{Boundary conditions at radial surfaces:} 

The radial \( (r_c)\) boundary condition can be enforced by adding a term to the loss function that penalizes the deviation from zero of the sum of the partial derivatives with respect to \( x \) and \( z \) at \( r = r_c \). The loss function term can be written as
\begin{equation}
\mathcal{L}_{\text{BCr}}^{M} = \frac{1}{\kappa_p}\text{MAE} \left( \frac{\partial p}{\partial x} \Big|_{r = r_c} + \frac{\partial p}{\partial z} \Big|_{r = r_c} \right) , \text{  } \in {\Omega}_{M}
\label{eq:Lmbcr}
\end{equation}

\subparagraph{Pre-training:} 
In the pretraining stage, we assumed that the matrix collocation points corresponding to the fractured zones are acting as the boundary condiitons of the system, following the pressure and saturation trends of the fractures. The matrix and fracture points, however, was calculated via Buckley-Leverett technique. The

\subsection{Flow in Fracture Domain}

Same as the matrix domain, we define the fracture total loss term, \(\mathcal{L}_{t}^{F} \) .

\subparagraph{PDE residuals:} The conservation of mass in fracture media should also be solved in parallel with the matrix equations. By utilizing equation (\ref{eq:massbalF}), we write

\begin{align}
{{\mathcal{L}}}_{w}^{F} = \frac{1}{\kappa_r} \text{MAE}(\text{  }\phi^ \parallel {\partial }_{t}\left({\rho}_{w}{s}_{w}^{F}\right) + \nabla ^ \parallel \left({\rho}_{w}{u}_{w}^{ \parallel }\right)~ + ~\frac{{\rho}_{w}}{{e}_{V}}\left(2{v}_{w}^{\bot }\right)\text{  }),\text{  } \in {\Omega}_{F}
\label{eq:Lfw}
\end{align}
\begin{align}
{{\mathcal{L}}}_{nw}^{F} = \frac{1}{\kappa_r} \text{MAE}(\text{  }\phi^ \parallel {\partial }_{t}\left({\rho}_{nw}{s}_{nw}^{F}\right) + \nabla ^ \parallel \left({\rho}_{nw}{u}_{nw}^{ \parallel }\right)~ + ~\frac{{\rho}_{w}}{{e}_{V}}\left(2{v}_{nw}^{\bot }\right)\text{  }),\text{  } \in {\Omega}_{F}
\label{eq:Lfnw}
\end{align}

Our approach to address the reported limitations of PINNs in capturing shock fronts for fully-viscous problems \citep{Fuks2020LIMITATIONSMEDIA} involved incorporating a small capillary pressure term. This term introduces a diffusive effect that mitigates the overly sharp transitions typically observed in the displacement front of the replacing fluid. Then, the total loss value for the residual of PDE in the fracture domain is calculated by
\begin{align}
{{\mathcal{L}}}_{PI}^{F} = ( {\mathcal{L}}_{w}^{F}  + {\mathcal{L}}_{nw}^{F} ) \cdot  e^{\omega^F}, \text{  } \text{  } \in {\Omega}_{F} 
\label{eq:Fvolbalance}
\end{align}
Here, \( e^{\omega^F}\) is defined as a local self-adaptive weight modifier (see Appendix \ref{sec:PINNsComputationalstrat}). A loss term is defined to constrain the values of the self-adaptive modifier
\begin{align}
{{\mathcal{L}}}_{\omega}^{F} = \text{MAE}(\omega^F , 0.0),\text{  } \in {\Omega}_{F}
\label{eq:»LomegaF}
\end{align}

\subparagraph{Initial conditions:}

A similar initial condition is also applied to the fracture collocation points.
\begin{align}
{{\mathcal{L}}}_{IC,s_w}^{F} = \text{MAE}({s}_{w}^{F}(x,y,z,t = 0) , {s}_{wc}),\text{  } \in {\Omega}_{F}
\label{eq:LeqICpfracture}
\end{align}
\begin{align}
{{\mathcal{L}}}_{IC,p_{nw}}^{F} = \frac{1}{\kappa_p}\text{MAE}({p}_{nw}^{F}(x,y,z,t = 0) , {p}_{i}),\text{  } \in {\Omega}_{F}
\label{eq:LeqICpfracturep}
\end{align}

\subparagraph{Boundary conditions at inlet:}

For the fractures with collocation points at the inlet boundary, we may also define a similar set of boundary conditions 
\begin{align}
{\mathcal{L}}_{BC0F,s_w}^{F} = \text{MAE}({s}_{w}^{F}(x,y = 0,z,t),{s}_{w,max}),\text{  } \in {\Omega}_{F}
\label{eq:LBC}
\end{align}

\begin{align}
{\mathcal{L}}_{BC0F,p_{nw}}^{F} = \frac{1}{\kappa_p}\text{MAE}({p}_{nw}^{F}(x,y = 0,z,t),{p}_{in}),\text{  } \in {\Omega}_{F}
\label{eq:LBCp}
\end{align}

\subparagraph{Boundary conditions at outlet:}

Similarly, for fractures with collocation points at the outlet boundary, we define a similar pressure boundary condition to satisfy the downstream pressure of the system in the fracture system.
\begin{align}
{{\mathcal{L}}}_{BC1,pnw}^{F} = \frac{1}{\kappa_p}\text{MAE}({p}_{nw}^{F}(x,y = L,z,t),{p}_{out}),\text{  } \in {\Omega}_{F}
\label{eq:LpnwBC1F}
\end{align}

\subsection{Observations}
During the inverse calculations, ensuring the PINNs predictions closely align with experimental observations is crucial. To achieve this, we define a data loss term, denoted by \(\mathcal{L}_{t}^{D} \), which represents the sum of one or more individual loss terms detailed below:

\subparagraph{Recovery factor (RF):}
As matching the average recovered fluids at all times is time-consuming and computationally intensive, we matched the RF at one randomly chosen time point during each epoch. This approach helps the network learn from the observation data without imposing a high computational burden on the optimizer. So, consider \( t_r \) as the time corresponding to one of the available CT-scan images, selected randomly, we can define \(  \bar{s_w} \) as:
\begin{align}
RF = \frac{1}{N_x N_y N_z} \sum_{i=1}^{N_x} \sum_{j=1}^{N_y} \sum_{k=1}^{N_z} s_w(xi, y_j, z_k, t=t_r), \text{  }  \in {\Omega}_{M}
\label{eq:LRF}
\end{align}

In the above equation, it is assumed that the core is fully-saturated by the non-wetting phase at the beginning of the test. Then, the corresponding loss term will be:

\begin{align}
{{\mathcal{L}}}_{{RF}} = \text{MAE}( RF , \bar{s_w}^{obs}(t=t_r) ), \text{  }  \in {\Omega}_{M}
\label{eq:LRF}
\end{align}

Here,  $\bar{s_w}^{obs}$  is the observed $\bar{s_w}$ at time \( t_r\), obtained by averaging the water saturation from the CT-scan images.

\subparagraph{In-situ water saturation:} To incorporate the measured in situ saturation data, a loss term is defined that minimises errors in local saturation values:

\begin{align}
\mathcal{L}_{s_w} = \text{MAE}( s_w(x,y,z,t), s_{w}^{\text{obs}}(x,y,z,t) ), \quad \text{for} \quad (x,y,z,t) \in \Omega_{M}
\label{eq:LRF}
\end{align}

where \( s_{w}^{\text{obs}} \) is the measured in-situ water saturation data. Given access to approximately one million observation points in the spatiotemporal dimensions, we randomly selected a small fraction of them (25000 points) at each epoch. In this work, the selected points was resampled (randomly) at each epoch.

\subparagraph{Inflow rate constraint:}
As an observational data, we may have access to the total inflow rate versus time. To match the inflow rate with the observational data, we write it as:

\begin{align}
{{\mathcal{L}}}_{{Q}_{inj}} = \text{MAE}( A_m \underset{}{\overset{}{\iint }}{u}_{w}{M}(y = 0)dxdy\text{  } + ... \notag \\
A_f \underset{}{\overset{}{\iint }}{u}_{w}^{F}(y^{F} = 0)dx^{F}dy^{F}\text{  },\text{  }{Q}_{inj}^{obs}(t)\text{  }) , \text{  }  \in ({\Omega}_{M} \cup {\Omega}_{F})
\label{eq:Lqwinj}
\end{align}

We denote the cross-sectional areas of the matrix and fracture as $A_m$ and $A_f$, respectively. To estimate $A_f$, we sum the areas of all fracture collocation points at the $y = 0$ plane, excluding overlaps. The area of each individual fracture collocation point is calculated using $\pi \left( \frac{e_v}{2} \right)^2$, where $e_v$ represents the fracture aperture. Finally, we determine $A_m$ based on the core radius ($r_c$) and the calculated $A_f$ using the formula $A_m = \pi(r_c)^2 - A_f$.

\section{Experimental Data/Collocation Points }
\label{sec:PINNsCollocation Points}

\subsection{Experimental Data}

The collocation points corresponding to the matrix and fractures are extracted from the clinical CT and micro-CT images . Each snapshot of the CT-scan had dimensions of [120, 87, 120], with the voxel resolution of 0.024 mm\textsuperscript{3}. We had access to 19 snapshots through the text from 0.1 to 311.405 hrs after starting of the injection. 
In total, we identified and extracted four fractures from the CT-scan observations, ensuring none of them intersected with each other. The spatial geometry of the fractures is shown in Fig. \ref{fig:coreschematic}b. We neglected the impacts of micro-fractures, although they were observed in the micro-CT scan images.

\subparagraph{Collocation points:} 
We needed to collect collocation points separately for matrix and fracture domains. The spatial coordinates of the collocation points were extracted on the basis of the coordinates of the points in the CT-scan images. The matrix spatial points have been chosen with the resolution of 30 points in  \(x\) and \(z\) (\(dx=dz\)= 0.81 mm) dimensions, and 60 points in  \(y\) dimension (\(dy\)= 0.96 mm). The collocation points close to the fracture collocation points have been removed from the list of spatial points. In total, $N=$ 73,500 spatial points have been used in the cylindrical domain of the core. Refer to Fig. \ref{fig:colls} for a visualization of these collocation points.
To address the system's boundary conditions, we've also collected the requisite boundary collocation points.
Fig. \ref{fig:colls} shows the collected collocation points for the boundary conditions at y=0, y=L and r=\(r_{c}\). The radial collocation points have been chosen via  \(x_{r} = r_c.cos(\beta)\) and \(z_{r} = r_c.sin(\beta)\), , where \(\beta\) represents the angular coordinate of the point, randomly generated within the range of \( (0, 2\pi) \). The corresponding $y$ values of the points were extracted from the main collocation points. 

\subparagraph{Fracture collocation points:} 
The fracture collocation points, were manually extracted from microCT scans, along with porosity and saturation images. Although some of the fractures could be viewed as separate smaller fractures, we merged the fractures that aligned with each other, reflecting their consistent behaviour in passing the fluids during the injection process. In total, $N^F$=4103 spatial collocation points were chosen in the fracture domain.

\subparagraph{Temporal points:} 
The temporal values assigned to the collocation points were randomly selected and subsequently updated during the resampling process. Anticipating spontaneous imbibition as the dominant mechanism, we selected collocation points at equal intervals relative to their square root of time values, ensuring a balanced distribution across the temporal dimension. \cite{Abbasi2023SimulationNetworks} have previously shown that the spontaneous imbibition  occurs with a constant rate at the square root of time coordinate, and it is advantageous to distribute the collocation points accordingly during training of PINNs.

\section{PINNs Computational Strategies }
\label{sec:PINNsComputationalstrat}

\subparagraph*{Adaptive loss weights:  }
In the training of PINNs, especially in complex processes, simultaneous minimization of various terms often requires addressing conflicting objectives. Numerous studies have demonstrated that employing adaptive weights for these loss terms can enhance the performance of PINNs \citep{Xiang2022Self-adaptiveNetworks}. In this study, we applied these techniques by employing self-adaptive weight values for both the loss terms and the calculated PDE residuals at different domains:  matrix \((\Omega_M)\), fracture \((\Omega_F)\),  and matrix-fracture \((\Omega_{MF})\). This adaptive weighting strategy enables more effective training by dynamically adjusting the contribution of different terms, leading to more reliable convergence.

In the context of adaptive-weighting of the PDE residuals at domain collocation points, using trainable vectors was impractical due to the high number of collocation points and the random selection of the temporal points. Therefore, we  defined different MLP networks - with spatiotemporal vectors as inputs, and adaptive weights at outputs - to provide self-adaptive weights for PDE residuals at various collocation points, as suggested by \citet{Anagnostopoulos2023Residual-basedPINNs}:
\begin{align}
{\omega}^{M} = {\mathcal{N}_{\omega}^{M}}(x,y,z,t,\theta)
\label{eq:adaptive_loss_matrixM}
\end{align}
\begin{align}
{\omega}^{F} = {\mathcal{N}_{\omega}^{F}}(x_f,y_f,z_f,t_f,\theta)
\label{eq:adaptive_loss_matrixF}
\end{align}
\begin{align}
{\omega}^{MF} = {\mathcal{N}_{\omega}^{MF}}(x_f,y_f,z_f,t_f,\theta)
\label{eq:adaptive_loss_matrixMF}
\end{align}

The MLP networks had a depth of 4 layers, each with a width of 40 neurons, activated by \( tanh\) activation functions. The weights of these networks were updated during each backpropagation gradient descent step.

\subparagraph*{Resampling of collocation points: }
The efficient simulation of the processes in 4D spatio-temporal spaces requires a significant number of collocation points to be chosen and computed during the training. These calculations create a significant computational load that makes the computations time-consuming. Various strategies, including resampling, have been proposed to address this challenge \citep{Coulaud2023Physics-InformedTransfer}. 
In our study, we implemented a resampling strategy for the collocation points, introducing randomness at various unique periods during training. Specifically, resampling for the matrix collocation points occurred every 10 epochs, with approximately 33\% of the spatial points being selected for resampling each time. The uniform sampling strategy inherently leads to a set of collocation points with statistical homogeneity. We have not applied the resampling strategy for the spatial points of the fractures.

Furthermore, the temporal coordinates corresponding to the points of the matrix and fractures were randomly resampled using a non-uniform square root of time \((\sqrt t) \) distancing approach in the temporal range of [1 - 10$^6$] seconds. For the temporal points also the resampling period was 10 epochs.

\subparagraph*{Loss term (error) thresholding:  } To enhance the training efficiency and mitigate the risk of overfitting, we have incorporated an error thresholding technique into the PINNs model. This approach involves neglecting the impact of loss terms that fall below certain preset threshold values. By introducing these threshold limits, we effectively disregard the contributions of loss terms that are deemed insignificant or negligible, thereby simplifying the training landscape. 

\begin{align}
\text{e} = \max(0, e - \tau) \times w
\label{eq:threshholding}
\end{align}

In this equation, $e$ is the value of the loss term, and \(\tau \) is the threshold value. We applied this technique - with the value of $\tau=0.003$ - to modify the values of boundary and initial conditions for both saturation and pressure terms.

\subparagraph*{Mathematics of the inverse calculations: }
A critical strategy in inverse computations using PINNs is the pre-training stage, at the beginning of the calculations. In the stage, the inverse parameters will be frozen for some epochs - about 2000 epochs in this work. This technique allows the model to learn the basic mathematical features of the system, before looking for the values of the inverse parameters. The length of pre-training stage depends on the complexity of the evaluating process.

After the freezing stage concluded, the inverse parameter values were updated at the end of each epoch, at same time as the network trainable parameters.
The value of the inverse parameter \(\gamma\) have been calculated based on
\begin{align}
\gamma=\gamma_{i}e^{\kappa \theta^*}
\label{eq:inverse_eq}
\end{align}
where \(\gamma_{i}\) is the initial value of the inverse parameter. Also, \( \theta^* \) is the trainable parameter in neural networks, which have been initialised to be zero at the beginning of the training. \( \kappa \) is a regularisation term that is responsible for the training rate of the parameter. We have found that there should be a balance between the model training speed and the inverse parameter. The large values of \( \kappa \) makes the calculations unstable and prone to converge to the wrong values. The exponential form of the equation allows the algorithm to vary in scales of magnitude. Also, it prevents the model to reach to the nonphysical negative values. 

To consider the uncertainties in the calculations, we defined a strategy to initialize the \(\gamma_{i}\) values randomly. It helps us in reaching a wide range of possible solutions for the problem, not only finding same solutions. Considering an initial value for the parameter of interest as \(\gamma_{i}^{*}\), we write
\begin{align}
\gamma_{i} = \gamma_{i}^{*} * 10^{\xi\xi_m}
\label{eq:inverse_eq}
\end{align}

A random number, denoted by $\xi$, is drawn from a uniform distribution between -1 and 1. This randomness is scaled by a model hyperparameter, $\xi_m$, which controls the overall level of randomness injected into the system. In this case, $\xi_m$ was set to 0.6.

\begin{figure}[htbp]
\begin{center}
\includegraphics[width=.9\textwidth]%
    {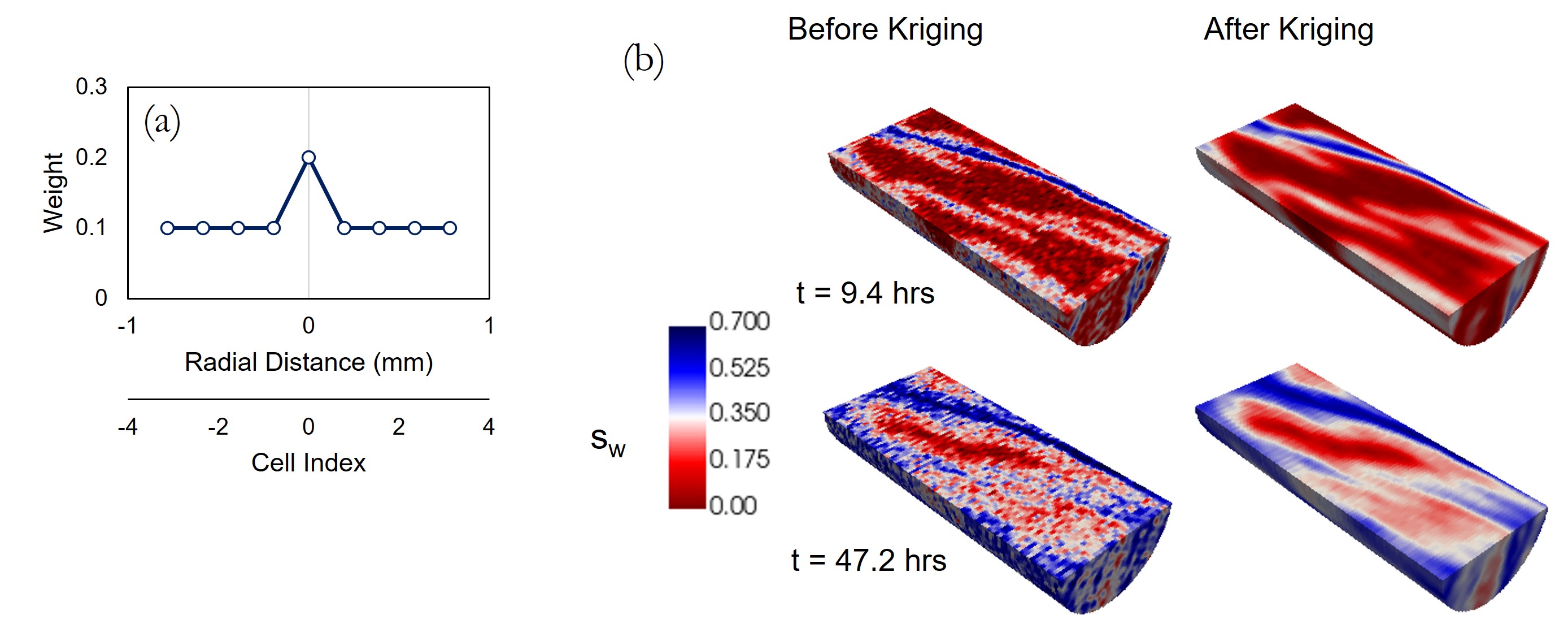}
\end{center}
\caption{The impact of 3D convolutional kriging on the in-situ water saturation data. a) The applied weighting kernel, b) A comparisons of the saturation distribution before and after kriging at two different times}
\label{fig:convkrig}
\end{figure}

\subparagraph*{Convolutional kriging: }
To overcome with the measurement noises in the measured in-situ saturation data, we applied 3D denoising operations on the dataset. We employed a 3D convolutional kriging approach to effectively interpolate between the cell properties and their neighbour property values. Convolutional denoising involves convolving a noisy data with a weighting kernel that performs smoothing to reduce the noises in the observations, while preserving the main features of the data. The kernel is an important part of the denoising. In this work, we utilized a semi-Gaussian weighted kernel that is shown in Fig. \ref{fig:convkrig}a. The algorithm of the computations is shown in Algorithm \ref{alg:convkrig}. We have used similar kernels for all three dimensions. Finally, a comparison of the saturation distribution in the rock after some time is shown at two different experiment times is shown in \ref{fig:convkrig}b. We can see that the approach could effectively capture the main features of the data while removing unnecessary noise.
 
\begin{algorithm}
\small 
\caption{3D Convolutional Kriging}
\begin{algorithmic}[1]
\REQUIRE $inp$: 3D array of input data
\STATE $k1, k2, k3 \leftarrow \begin{pmatrix} 1 & 1 & 1 & 1 & 2 & 1 & 1 & 1 & 1 \end{pmatrix}$ \COMMENT{the kernel along the dimensions}
\STATE $k1, k2, k3 \leftarrow \text{normalize}(k1, k2, k3)$ \COMMENT{Normalize the kernels to ensure the mass balance is attained.}
\STATE $out \leftarrow inp$ \COMMENT{Initialize output array as input}
\FOR{$i$ in range($0$, $3$)}
    \STATE $out \leftarrow \text{convolve}(out, k_i, \text{axis}=i)$ \COMMENT{Convolve over dimension $i$ }
\ENDFOR
\RETURN $out$
\end{algorithmic}
\label{alg:convkrig}
\end{algorithm}

\subparagraph*{Normalization and denormalization: }
An essential pre/post processing step in training of PINNs is normalization and then denormalization of the inlet and outlet variables. By standardizing the input/output data into a common scale and distribution, it can greatly enhance the model's stability and performance. In this work, we have utilized a Z-score normalization scheme. We normalized the PINNs inlets as

\begin{align}
x_n = \frac{{x - \bar{x}}}{{\sigma}}
\label{eq:norm}
\end{align}
and then the outlets denormalized via
\begin{align}
x = x_n \cdot \sigma + \bar{x}
\label{eq:norm}
\end{align}
where,  $\bar{x}$ and $ \sigma$ are the mean and standard deviation of the variable \(x\) in the domain under study, respectively.

\subparagraph*{Fast Fourier transformation (FFT): } In previous studies, various works have highlighted the challenges of PINNs in solving high-frequency solutions due to an issue known as spectral bias \cite{Fridovich-Keil2021SpectralGeneralization}. This bias causes the network to preferentially learn solutions with lower frequencies, not necessarily the most true solutions. A prominent strategy to address this involves incorporating Fast Fourier Transformation (FFT) within the network architecture \cite{Wang2023AnNetworks,Abbasi2023SimulationNetworks,Wang2021OnNetworks}. In this work, we also utilized a FFT operator, same as \citet{Abbasi2023SimulationNetworks} ,to make the network able to capture the high-frequency characteristics of the solution.  As it is visualized in Fig. \ref{fig:2}, the normalized inlets are first encoded, then transformed to the frequency domain using an FFT operation.  After processing by the main Multi-Layer Perceptron (MLP) network, the information is returned to the spatial domain via an inverse FFT operator. Only the real parts of the solution are retained, discarding the imaginary components.  Table \ref{table:fft} compares the performance of the workflow for solving the benchmark model, when different strategies of FFT transformation was applied. It confirms the effectiveness of the applied strategy in improving the performance of the model.

\begin{table}[ht]

\caption{The impact of using different Fourier Transformation strategies on the accuracy of the final results for the forward solution of the benchmark problem}
\begin{center}
    \small 
    \begin{tabular}[t]{lll}
    \hline
    Case & Lt (MAE) & RF (MAE)   \\ \hline
    without FT           &  2.51e-3 & 6.6e-2  \\ \hline
    with FT (imaginary)  &  5.30e-4 & 2.4e-2  \\ \hline
    with FT (real)       &  4.43e-4 & 2.2e-2  \\ \hline
  
    \end{tabular}
\end{center}
\label{table:fft}
\end{table}

\section{Buckley-Leverett equation}
\label{sec:BLequation}

At the first stage of the PINNs calculations, we have assumed the flow in fractures is governed by the Buckley-Leverett assumptions, meaning that flow is assumed to be a sole function of viscous forces, while capillary and gravity forces are neglected.  

As a basic solution to this problem, we may follow the solution recommended by \citet{buckley1942mechanism}, called the Buckley-Leverett (BL) equation. The solution assumes that two fluids are incompressible and that the flow occurs unidirectionally.  Rock properties (e.g., porosity and permeability) is considered homogeneous and constant. 
Also, the problem is considered isothermal with no interphase mass transfer between fluids. The mass conservation equation may be written as  \citep{Lake1989EnhancedRecovery}: 
\begin{equation}
\twoPflow
\end{equation}

Rewriting the equation for a 1D problem, and assuming incompressibility of rock and fluids, leads to a further simplification as: 
\begin{equation}
\oneDtwoP
\end{equation}
where, \(u_i\) is the Darcy flow velocity of each phase and is determined by Darcy's law:

\begin{equation}
\darcy
\label{darcy}
\end{equation}

Equation (\ref{darcy}) can be rewritten by eliminating pressure in the equation and summing the phase equations of two phases:

\begin{equation}\label{oneDfractionalform}
\oneDfractionalform
\end{equation}
where, \(f_i\) is the fractional flow of phase water, and is written as:
\begin{equation}\label{oneDfractionalform}
\fractionalflow
\end{equation}

We then employ the Welge tangent method \citep{Welge1952ADrive} to calculate the local water saturation at each time step, as shown in the following equation
\begin{align}
x = \frac{u_t}{\phi} \frac{d f_w}{d s_w} t 
\label{eq:norm}
\end{align}
where, \( u_t\) is the total velocity of the fluids, in fracture media at this case, which we can calculate via
\begin{align}
u_t = \frac{K^F {k}_{rw}^{max}}{\mu_w}  \frac{(p_{in}-p_{out})}{l}  
\label{eq:norm}
\end{align}

\section{Appendix: Computational Resources }
\label{sec:compres}
In this study, we compared two simulation approaches: Finite Difference (FD) and PINNs. We utilized an in-house FD-based numerical simulator \cite{Lohne2017ARegimes} as a benchmark tool. The FD numerical simulator were developed in C++ and run on an Intel i9-11950H CPU (2.60 GHz) with 32 GB RAM. Individual forward simulations for a 27,000 cell model ranged from 30 minutes to 2 hours, depending on the complexity of effective parameters (e.g., flow parameter shapes and matrix-fracture property contrasts). The simulator were designed to effectively capture the experimental boundary conditions. 

In contrast, PINNs computations were conducted using a single NVIDIA RTX A2000 GPU with the PyTorch 1.13.1+cu117 framework. Notably, PINNs simulations exhibited almost similar runtimes for both forward and inverse problems. For a standard inverse computation scenario, with a typical network size and number of collocation points as reported in Table \ref{tab:2}, the inverse computations typically required nearly 1.5 - 2.5 hours.

\section{Appendix: History Matching }
\label{sec:historymatching}
The PINNs history-matching computations were performed using Adam optimizer on the PyTorch computational framework. To compare with PINNs, we coupled the Finite-Difference (FD) simulator with a Nelder-Mead (NM) simplex optimization algorithm \cite{Nelder1965AMinimization}. The optimizor was deployed in SciPy library \cite{Virtanen2020SciPyPython}. We used the default meta-parameters.

\begin{figure}[htbp]
\begin{center}
\includegraphics[width=.99\textwidth]%
    {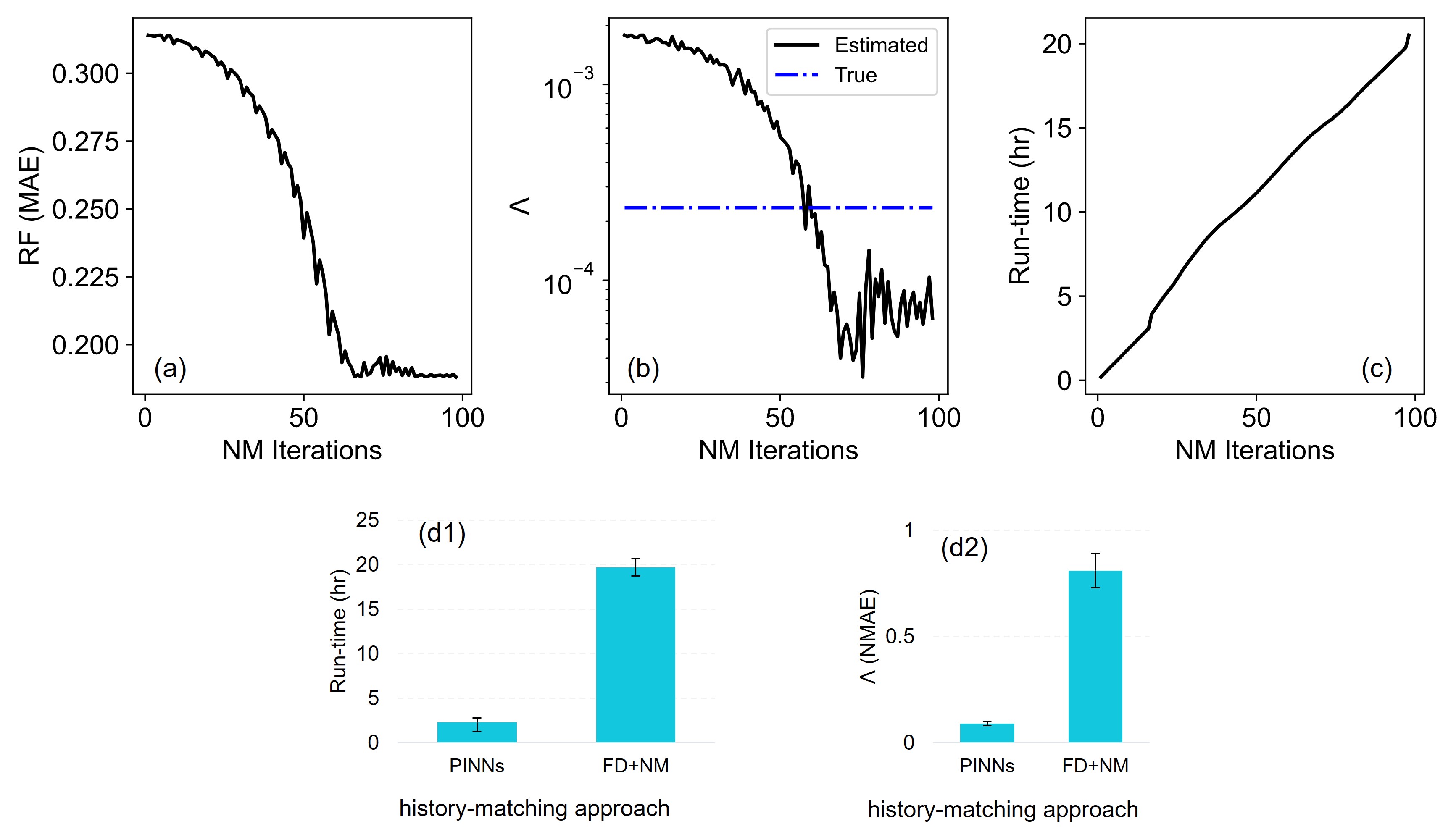}
\end{center}
\caption{\textbf{The results of history-matching using coupled finite-difference numerical simulations, and Nelder-Mead (NM) optimization.} The history-matching continued for around 100 NM iterations. a) The MAE in the matched RF, b) The estimated $\bar{\Lambda}$ curve compare to the expected true value, c) The total run-time versus NM iterations, d) the performance (1: run-time, 2: NMAE in the estimations) of PINNs, and FD-NM history-matching tools are compared.}
\label{fig:FDNM}
\end{figure}

For the synthetic problem, the static model of the system kept constant for all the cases. The goal was only history-matching of the RF curve. To achieve faster results while accepting a certain level of error, we utilized a static model with 15,625 cells, which is less than the original synthetic model. A summary of the optimization is shown in Fig. \ref{fig:FDNM}. After approximately 100 iterations, the errors in the objective function stabilized. As seen in the figure, the history-matching computations took around 20 hours to stabilize. Even after stabilization, the error in the estimated $\Lambda$ curve was significant, with a normalized mean absolute error (NMAE) of 0.81. Fig. \ref{fig:FDNM}d shows the clear advantage of PINNs-based history-matching compared to FD-NM history-matching tool.

\section{Appendix: Sensitivity Analysis }
\label{sec:sensitivityanalysis}

\begin{figure}[htbp]
\begin{center}
\includegraphics[width=.99\textwidth]%
    {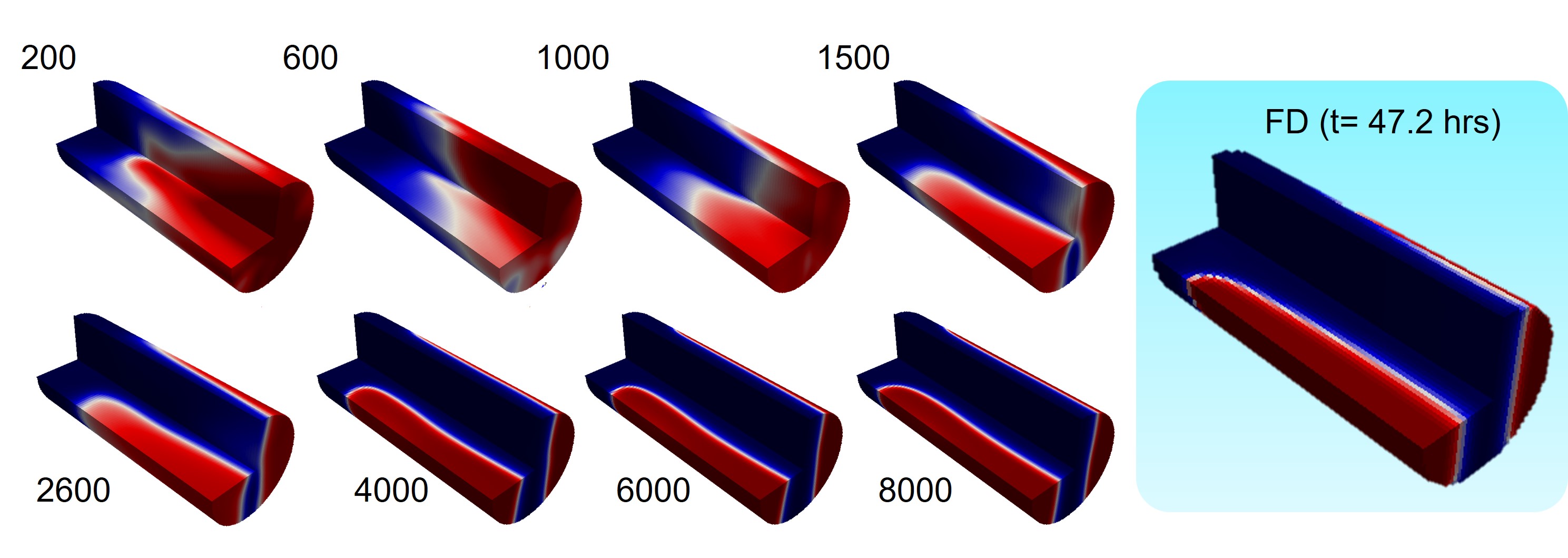}
\end{center}
\caption{The dynamics of training PINNs at different epochs (200-8000) compared to the FD simulation result at same time (t=47.2 hrs).}
\label{fig:b_trdyna}
\end{figure}

\subparagraph*{Training dynamics. }
The dynamics of training of PINNs is visualized in Fig. \ref{fig:b_trdyna}. It is evident that the precision of PINNs computations is enhanced as the training advances. Initially, the PINNs predictions encapsulate the high-dimensional attributes of the flow within the matrix. Following this, the interactions between matrix and fracture flows have been learnt.

\begin{figure}[htbp]
\begin{center}
\includegraphics[width=.7\textwidth]%
    {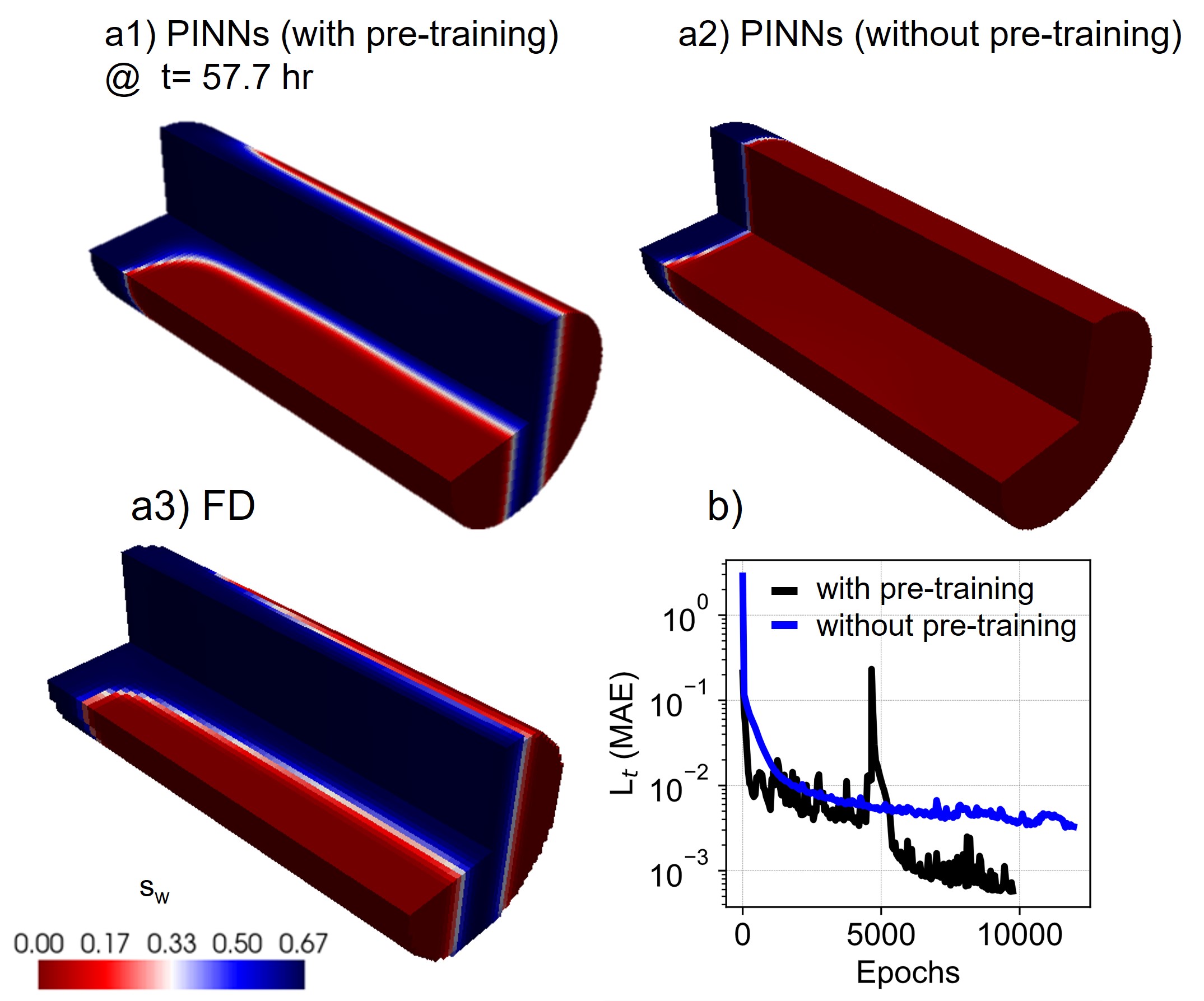}
\end{center}
\caption{The impact of using pre-training strategy on the accuracy of the final solutions. a) the saturation distributions compared, b) a comparison of value of total loss function through the training. }
\label{fig:pretraining}
\end{figure}

\subparagraph*{Impact of pre-training. }
To show how performing a pre-training strategy is vital in reaching the final solution of the system, in this section, we compare the solutions in which the training has been performed without pre-training. As shown in Fig. \ref{fig:pretraining}, the case without pre-training could not capture the high-frequency component of the solution, that is, the impacts of fractures on the saturation distribution. This could be related to spectral bias, where the network tends to capture only low-frequency components of the solution. In the pre-training stage, also called curriculum learning, by step-wise increasing the complexity, helps the PINNs model to avoid spectral biases \cite{Toscano2024FromLearning}.

\begin{figure}[htbp]
\begin{center}
\includegraphics[width=.9\textwidth]%
    {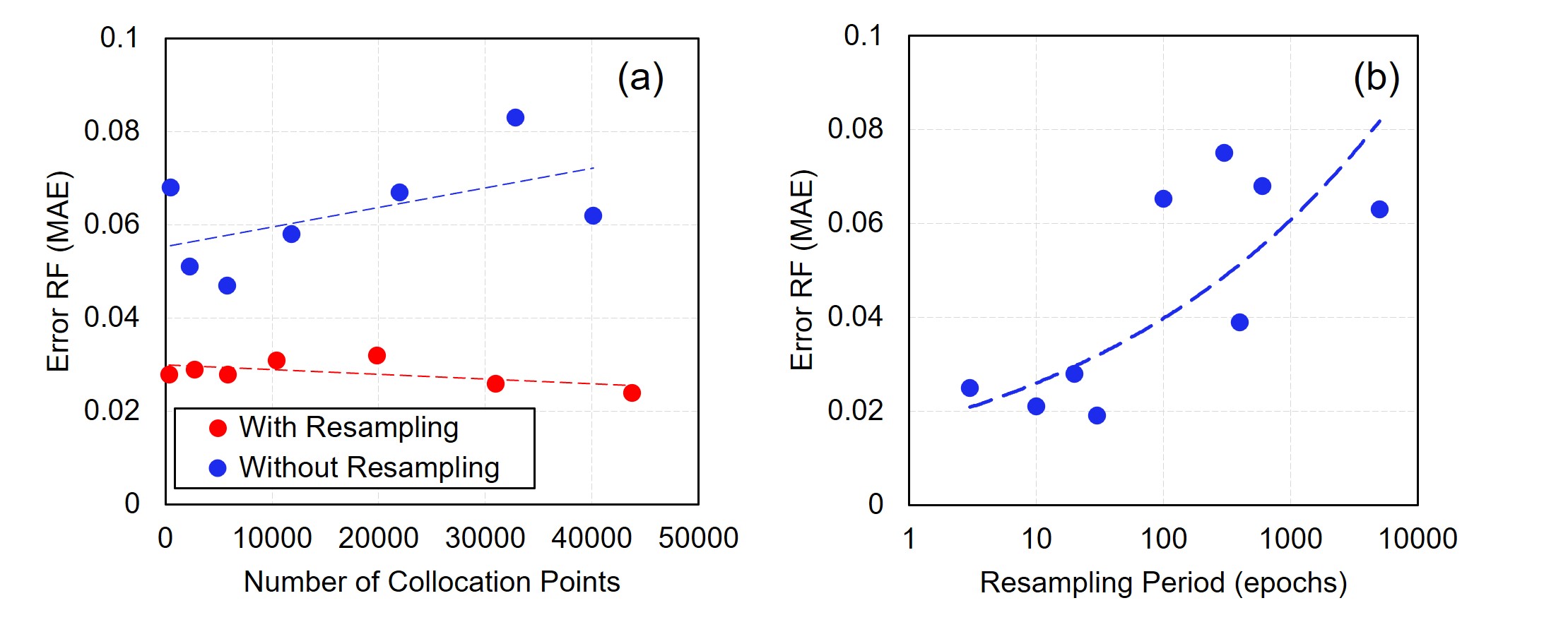}
\end{center}
\caption{The sensitivity analysis on the number of collocation points in the matrix. All the models were trained for 20000 epochs. a) A comparison of the impact of the number of collocation points on the MAE in the predicted RF curve, b) The impact of resampling period on the accuracy of the predictions.}
\label{fig:reselectioncoll}
\end{figure}

\subparagraph*{Resampling of collocation points. }  As it is demonstrated in Fig. ~\ref{fig:reselectioncoll}, utilizing resampling techniques significantly reduces errors while accelerating the computations. Resampling, applicable for both spatial and temporal domains, helps the network avoid getting stuck on local regions and steers it towards the optimal general solution of the system in the domain. This is particularly advantageous for large-scale problems that require a substantial number of collocation points. 

As Fig.~\ref{fig:reselectioncoll}b illustrates, the effectiveness of the resampling strategy depends on the frequency of resampling, or how often it's applied during training epochs. Resampling with very low frequency can not significantly help the optimization. Conversely, excessively frequent resampling prevents the system from learning the selected points effectively. In this context, a frequency between 0.03 and 0.1 (corresponding to resampling every 10-30 epochs) appears to be the optimal choice.

\end{document}